\documentclass[10pt,conference]{IEEEtran}
\usepackage{cite}
\usepackage{amsmath,amssymb,amsfonts}
\usepackage{graphicx}
\usepackage{textcomp}
\usepackage{xcolor}
\usepackage[hyphens]{url}
\usepackage{fancyhdr}
\usepackage{hyperref}
\usepackage{balance}
\usepackage{comment}

\usepackage[]{hyperref}

\usepackage{enumitem}
\usepackage{algorithm}
\usepackage[noend]{algpseudocode}
\usepackage{makecell}
\usepackage{multicol}
\usepackage{multirow}
\usepackage{tikz}

\definecolor{ForestGreen}{RGB}{34,139,34}

\newcommand{\RN}[1]{%
  \textup{\uppercase\expandafter{\romannumeral#1}}%
}
\newcommand*\circled[1]{\tikz[baseline=(char.base)]{\node[shape=circle,draw,inner sep=0.85pt, fill=black] (char) {#1};}}

\newcommand{\mcf}{{\fontfamily{lmss}\selectfont
mcf}}
\newcommand{\mcfs}{{\fontfamily{lmss}\selectfont
mcf}}

\newcommand{\lbm}{{\fontfamily{lmss}\selectfont
lbm}}
\newcommand{\lbms}{{\fontfamily{lmss}\selectfont
lbm}}

\newcommand{\pr}{{\fontfamily{lmss}\selectfont
PageRank}}
\newcommand{\prs}{{\fontfamily{lmss}\selectfont
pr}}

\newcommand{\motif}{{\fontfamily{lmss}\selectfont
Motif Mining}}
\newcommand{\motifs}{{\fontfamily{lmss}\selectfont
motif}}

\newcommand{\rmone}{{\fontfamily{lmss}\selectfont
DLRM(MemBound)}}
\newcommand{\rmones}{{\fontfamily{lmss}\selectfont
rm1}}

\newcommand{\rmtwo}{{\fontfamily{lmss}\selectfont
DLRM(Balanced)}}
\newcommand{\rmtwos}{{\fontfamily{lmss}\selectfont
rm2}}

\newcommand{\llm}{{\fontfamily{lmss}\selectfont
Large Lang. Model}}
\newcommand{\llms}{{\fontfamily{lmss}\selectfont
llm}}

\newcommand{\redis}{{\fontfamily{lmss}\selectfont
Redis}}
\newcommand{\rediss}{{\fontfamily{lmss}\selectfont
redis}}

\newcommand{\stream}{{\fontfamily{lmss}\selectfont
Streaming}}
\newcommand{\streams}{{\fontfamily{lmss}\selectfont
stm}}

\newcommand{\rdom}{{\fontfamily{lmss}\selectfont
Random}}
\newcommand{\rdoms}{{\fontfamily{lmss}\selectfont
rand}}

\newlength\myindent
\newcommand{\THISWORK}{{\fontfamily{lmss}\selectfont
Palermo}}

\newcommand{\hpcayear}{2025}

\newcommand{\hpcasubmissionnumber}{37}
\title{Guidelines for Submission to HPCA \hpcayear{}}

\def\hpcacameraready{} 

\newcommand\hpcaauthors{Haojie Ye$\dagger$, Yuchen Xia$\dagger$, Yuhan Chen$\dagger$, Kuan-Yu Chen$\dagger$, Yichao Yuan$\dagger$, Shuwen Deng$\ddagger$,\\Baris Kasikci$^*$, Trevor Mudge$\dagger$, Nishil Talati$\dagger$}
\newcommand\hpcaaffiliation{$\dagger$University of Michigan, USA; $\ddagger$Tsinghua University, China; $^*$University of Washington, USA}
\newcommand\hpcaemail{Email: yehaojie@umich.edu}

\author{
  \ifdefined\hpcacameraready
    \IEEEauthorblockN{\hpcaauthors{}}
      \IEEEauthorblockA{
        \hpcaaffiliation{} \\
        \hpcaemail{}
      }
  \else
    \IEEEauthorblockN{\normalsize{HPCA \hpcayear{} Submission
      \textbf{\#\hpcasubmissionnumber{}}} \\
      \IEEEauthorblockA{
        Confidential Draft \\
        Do NOT Distribute!!
      }
    }
  \fi 
}

\fancypagestyle{camerareadyfirstpage}{%
  \fancyhead{}
  
  \fancyhead[C]{
    \ifdefined\aeopen
    \parbox[][12mm][t]{13.5cm}{\hpcayear{} IEEE International Symposium on High-Performance Computer Architecture (HPCA)}    
    \else
      \ifdefined\aereviewed
      \parbox[][12mm][t]{13.5cm}{\hpcayear{} IEEE International Symposium on High-Performance Computer Architecture (HPCA)}
      \else
      \ifdefined\aereproduced
      \parbox[][12mm][t]{13.5cm}{\hpcayear{} IEEE International Symposium on High-Performance Computer Architecture (HPCA)}
      \else
      \parbox[][0mm][t]{13.5cm}{\hpcayear{} IEEE International Symposium on High-Performance Computer Architecture (HPCA)}
    \fi 
    \fi 
    \fi 
    \ifdefined\aeopen 
      \includegraphics[width=12mm,height=12mm]{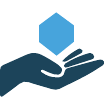}
    \fi 
    \ifdefined\aereviewed
      \includegraphics[width=12mm,height=12mm]{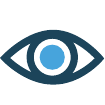}
    \fi 
    \ifdefined\aereproduced
      \includegraphics[width=12mm,height=12mm]{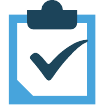}
    \fi
  }
  \fancyfoot[C]{}
}
\begin{document}

\title{\THISWORK: Improving the Performance of Oblivious Memory using Protocol-Hardware Co-Design}
\maketitle

\ifdefined\hpcacameraready 
  \thispagestyle{camerareadyfirstpage}
  \pagestyle{empty}
\else
  \thispagestyle{plain}
  \pagestyle{plain}
\fi

\newcommand{\hpcaheight}{0mm}
\ifdefined\eaopen
\renewcommand{\hpcaheight}{12mm}
\fi


\begin{abstract}

Oblivious RAM (ORAM) hides the memory access patterns, enhancing data privacy by preventing attackers from discovering sensitive information based on the sequence of memory accesses.
The performance of ORAM is often limited by its inherent trade-off between security and efficiency, as concealing memory access patterns imposes significant computational and memory overhead.
\textcolor{black}{While prior works focus on improving the ORAM performance by prefetching and eliminating ORAM requests, we find that their performance is very sensitive to workload locality behavior and incurs additional management overhead caused by the ORAM stash pressure.}

\textcolor{black}{This paper presents \THISWORK: a protocol-hardware co-design to improve ORAM performance.}
The key observation in \THISWORK\ is that classical ORAM protocols enforce restrictive dependencies between memory operations that result in low memory bandwidth utilization.
\THISWORK\ introduces a new protocol that overlaps large portions of memory operations, within a single and between multiple ORAM requests, without breaking correctness and security guarantees.
Subsequently, we propose an ORAM controller architecture that executes the proposed protocol to service ORAM requests.
The hardware is responsible for concurrently issuing memory requests as well as imposing the necessary dependencies to ensure a consistent view of the ORAM tree across requests.
Using a rich workload mix, we demonstrate that \THISWORK\ outperforms the RingORAM baseline by 2.8$\times$, on average, incurring a negligible area overhead of 5.78mm$^2$ (less than 2\% in 12th generation Intel CPU after technology scaling) and 2.14W without sacrificing security.
We further show that \THISWORK\ also outperforms the state-of-the-art works PageORAM, PrORAM, and IR-ORAM.

\end{abstract}

\section{Introduction} \label{section:introduction}

Building a secure oblivious RAM (ORAM) design for cloud workloads is critical because it ensures the confidentiality of accessing sensitive data. 
ORAMs are designed to offer a secure memory access pattern, concealing the information being accessed and safeguarding the privacy of sensitive data~\cite{goldreich1996software}.
This is particularly important in the context of modern AI workloads (\textit{e.g.,} recommendation and large language models), where the memory access pattern can reveal sensitive prior product selections, such as user preferences or prompts to the language model.
ORAM can protect against these potential vulnerabilities by \textit{obfuscating the access pattern to the memory}, preventing an attacker from inferring any useful information about where the data is being accessed.

However, the obfuscation of the memory access patterns using ORAM incurs a \textit{significant performance overhead}.
For example, to hide memory access in a 16GB protected memory space, classical ORAM implementations such as PathORAM~\cite{pathoram} and RingORAM~\cite{ringoram} on average convert a single access to 576 and 470 accesses, respectively.
This is because the critical metadata structure to protect the memory space \texttt{PosMap} is too large to fit on-chip. 
Addressing this issue necessitates a hierarchical ORAM design, often involving three levels, to establish a mapping between the physical and ORAM addresses~\cite{fletcher2015freecursive}.
Therefore, optimizing the performance of ORAM implementations is \textit{crucial for practicality}.

Prior works~\cite{che2020multi,iroram,proram,rajat2023laoram} present ways to optimize the performance of ORAM by prefetching and eliminating as many ORAM requests as possible. 
\textcolor{black}{However, we find two limitations of these works.
First, the benefit of these designs is highly sensitive to the application behavior, benefiting workloads with high spatial locality.
They fall short in optimizing other workloads with low data locality.
Second, prefetching-based optimizations modify the original PathORAM algorithm, which selects ORAM leaves independently and uniformly at random. 
These designs map consecutive physical addresses in the original memory space to the same leaf in the ORAM tree.
This practice significantly increases the ORAM stash pressure and introduces background evictions, ultimately resulting in diminishing returns and a capped speedup at 3.2$\times$ even with a synthetic trace exhibiting perfect locality.
}

This paper presents \THISWORK---a protocol-hardware co-design to improve ORAM performance.
The design objective of \THISWORK\ is to enhance performance without compromising prior ORAM security guarantees~\cite{pathoram}.
Using the performance analysis of RingORAM~\cite{ringoram}, we discover that the average memory bandwidth utilization of executing RingORAM on the tested benchmarks is less than 30\%.
Considering the memory-intensive nature of ORAM, this is surprising; our further analysis of reduced bandwidth utilization reveals that enforced dependencies in the protocol limit memory-level parallelism.

To this end, we design \THISWORK\ that re-architects the ORAM protocol, and designs a novel hardware ORAM controller architecture to fully unlock the potential of our protocol.
The protocol is designed to improve the concurrency in serving ORAM requests.
In particular, the \THISWORK\ protocol introduces intra- and inter-request parallelism to increase memory bandwidth utilization.
Intra-request parallelism involves overlapping memory requests related to different steps in accessing various hierarchical levels of ORAM.
Inter-request parallelism entails concurrent processing of memory requests across distinct ORAM requests.
While overlapping memory requests, the \THISWORK\ protocol identifies and enforces minimal dependencies,
ensuring a consistent view of the ORAM tree after each address remapping.

To support the proposed protocol, we also present the design of an ORAM controller.
The hardware architecture consists of on-chip memory structures for stash, position map, tree-top cache~\cite{maas2013phantom}, and a mesh of Processing Engines (PEs) that enable concurrent serving of ORAM memory requests.
\textcolor{black}{We present both qualitative and quantitative security analyses of \THISWORK\ to assess the uniform randomness of the attacker's view of the DRAM traffic and the isolation of each LLC miss latency, even when concurrently serving multiple ORAM memory requests at the ORAM controller.
}

To demonstrate the effectiveness of \THISWORK, we use a wide range of workloads including SPEC17~\cite{bucek2018spec}, graph analytics~\cite{beamer2015gap,mackey2018chronological}, deep learning~\cite{gupta2020architectural, gupta2020deeprecsys,radford2019language}, and key-value accesses~\cite{AmazonUserApi}.
\THISWORK\ outperforms PathORAM~\cite{pathoram} and RingORAM~\cite{ringoram} by 2.4$\times$ and 2.2$\times$, on average.
Borrowing ideas from prior prefetching works~\cite{che2020multi, proram, rajat2023laoram}, we show that \THISWORK\ can further improve this performance gain by 3.1$\times$ and 2.8$\times$, respectively.
\textcolor{black}{We also show improvements against PageORAM~\cite{pageoram}, PrORAM~\cite{proram}, and IR-ORAM~\cite{iroram} while upholding the same security level as in these works.}
The performance improvements are attributed to improved memory bandwidth utilization from 21\% in the RingORAM baseline to 59\% in \THISWORK.
\textcolor{black}{Our post-synthesis RTL results reveal that \THISWORK\ only consumes 5.78mm$^2$ silicon area (less than 2\% in Intel 12th Gen CPU after technology scaling~\cite{wikichip}, providing a low-complexity practical solution) and 2.14W.}
The key contributions of this work are as follows.
\begin{itemize}[leftmargin=*]
    \item Performance analysis of prior ORAM implementations revealing further optimization opportunities.
    \item Design of a new ORAM protocol that enables overlapping memory requests within the same ORAM request and between multiple ORAM requests.
    \item Design of the ORAM controller architecture to \textcolor{black}{apply the minimal required protocol synchronization and} improve memory-level parallelism.
    \item \THISWORK: end-to-end protocol-hardware co-design that improves an average performance of RingORAM by \textcolor{black}{2.8$\times$}.
\end{itemize}

\section{ORAM Background and Threat Model} \label{section:background}
\subsection{A Case For Oblivious RAM/Memory} \label{subsection:signal}
The evolution of cloud services has facilitated the execution of applications with exceedingly large memory and computing demands.
While clients enjoy the benefit of using these cloud services, clients may unintentionally give up the privacy of running their applications even when data is fully encrypted.
We briefly discuss an example of how an untrusted cloud can learn sensitive information from clients. 

Consider a client executing a Large Language Model (LLM) inference workload using untrusted outsourced memory to store the token feature table. 
The external memory party (attacker) has the capability to monitor the complete memory bus and eavesdrop on the memory request traces of the victim.
Even though the transferred feature values are fully encrypted, the attacker can still observe the full address request trace and infer about token address and frequency distribution. 
The attacker can iterate over common LLM models and map memory traces to corresponding tokens. 
In this way, the attacker can fully reconstruct all the user's prompts and extract sensitive knowledge from the user. 
The aforementioned example highlights various recent incidents and concerns related to corporate information leakage by ChatGPT~\cite{samsung, microsoft}. 
Such outcomes impose a substantial trust burden on the reliance on third-party cloud memory services.

This calls for an urgent need for privacy in using cloud memory services. 
One attractive privacy solution is to use the oblivious memory protocol (ORAM). 
ORAM enables the user to hide the memory access addresses, types, and patterns from the attacker when using the untrusted cloud memory service so that their access patterns are computationally indistinguishable from random accesses. 
\textcolor{black}{In fact, ORAM has been incorporated into Signal's ecosystem~\cite{signal}, facilitating users to conduct contact discovery while safeguarding the privacy of each user's social map through ORAM.
}

\subsection{Threat model}
\label{back:threat_model}
\textcolor{black}{
We model a common scenario that a shared server with cloud settings is launched to serve remote clients. 
The server is equipped with a standalone secure processor, \textit{i.e.,} the Trusted Computing Base (TCB), that includes cores and a small amount of on-chip memories. 
The cloud service can use any commodity off-chip memory modules and can snoop the memory bus as an attacker. 
The processor runs a private or public program on private data. 
On LLC misses, the processor issues to a trusted on-chip ORAM controller to access data in untrusted external memory. 
The attacker is ``honest but curious'' and uses any possible information to gather insights from the victim process, such as memory access hotspots, memory bus contents, timings, \textit{etc}. 
Any distinguishable behavior from the original LLC miss trace is considered an obliviousness violation~\cite{pathoram}.
This is a common threat model that is present in most prior ORAM works~\cite{rho,iroram,raoufi2023ab,proram, fletcher2015freecursive}. 
}

\subsection{ORAM Protocol}

\begin{algorithm}[t]
\scriptsize
\caption{Pseudocode for RingORAM algorithm to serve an ORAM request.}
\label{alg:form_pseudocode}
\begin{algorithmic}[1]
\Procedure{\textcolor{blue}{RingORAMAccess}}{\texttt{PA, op, data'}} \\
    \hspace*{\algorithmicindent} \textbf{Global}: \texttt{PosMap}: Position map that stores the mapped leaf ID for each PA \\
    {\textcolor{blue}{// PosMap stored in recursive ORAM due to linear size growth with DRAM capacity}}  \\
    \hspace*{\algorithmicindent} \textbf{Global}: \texttt{Stash}: Buffer that temporarily stores the data read from DRAM  \\
    \hspace*{\algorithmicindent} \textbf{Global}: \texttt{round}: Enforces a stash eviction for every A accesses 
    
    \State \texttt{leaf = PosMap[PA]}
    \State \texttt{leaf' = UniRandLeaf}
    \State \texttt{PosMap[PA] = leaf'}
    
    \State \texttt{data = ReadPath(leaf, PA)}

    \If{\texttt{op == READ}}
        \State \texttt{return data to processor from Stash}
    \Else
        \State \texttt{Stash[PA] = data'}
    \EndIf
    
    \If{\texttt{(round++) \% A == 0}}
        \State \texttt{EvictPath()}
    \EndIf

    \State \texttt{EarlyReshuffle(leaf)}
    
    \State \texttt{\textbf{return}}
\EndProcedure
\\
\Procedure{\textcolor{blue}{ReadPath}}{\texttt{leaf, PA}} \\
    \hspace*{\algorithmicindent} \textbf{Input}: \texttt{leaf}: leaf ID \\
    \hspace*{\algorithmicindent} \textbf{Input}: \texttt{PA}: Physical address that misses the LLC \\
    \hspace*{\algorithmicindent} \textbf{Global}: \texttt{NodeMetadata}: Keeps track of the ORAM tree per node metadata \\
    {\textcolor{blue}{// NodeMetadata stored in DRAM due to linear size growth with DRAM capacity}}  
    \ForAll{\texttt{NodeID} $\in$ \texttt{leaf along root}}
    \State \texttt{i = unused\_fake\_blk \textbf{if} PA in NodeID \textbf{else} real\_blk}
    \State \texttt{Stash = Stash $\cup$ ReadBucket(NodeID, i)}
    \State \texttt{NodeMetadata[NodeID].update()}
    \EndFor
\EndProcedure
\\
\Procedure{\textcolor{blue}{EarlyReshuffle}}{\texttt{leaf}} \\
    \hspace*{\algorithmicindent} \textbf{Input}: \texttt{leaf}: leaf ID
    \Comment{\textcolor{blue}{Check along leaf whether buckets can be further used}} 
    
    \ForAll{\texttt{NodeID} $\in$ \texttt{leaf along root}}
    \If{\texttt{NodeMetadata[NodeID].accessed == S}}
    \State \texttt{ResetBucket(NodeID)}
    \EndIf
    \EndFor
\EndProcedure
\\
\Procedure{\textcolor{blue}{EvictPath}}{\texttt{}} \\
    \hspace*{\algorithmicindent} \textbf{Global}: \texttt{G}: leaf ID to exercise eviction \\
    {\textcolor{blue}{// RingORAM uses a deterministic ring counter eviction leaf sequence}}  
    
    \ForAll{\texttt{NodeID} $\in$ \texttt{G along root}}
    \State \texttt{ResetBucket(NodeID)}
    \EndFor
    
    \State \texttt{G = next ring counter}
\EndProcedure
\\
\Procedure{\textcolor{blue}{ResetBucket}}{\texttt{NodeID}} \\
    \hspace*{\algorithmicindent} \textbf{Input}: \texttt{NodeID}: Node ID in the ORAM tree that needs a bucket reset

    \State \texttt{Fetch\_offset = NodeMetadata[NodeID].unused\_real\_blk()}
    \State \texttt{Fetch\_offset.pad(NodeMetadata[NodeID].unused\_fake\_blk())} \\
    {\textcolor{blue}{// Fetched offset size is padded to Z to ensure obliviousness on the memory bus}}
    \ForAll{\texttt{i} $\in$ \texttt{Fetch\_offset}}
    \State \texttt{Stash = Stash $\cup$ ReadBucket(NodeID, i)}
    \EndFor
    \State \texttt{WriteBucket(NodeID, Stash)}    \Comment{\textcolor{blue}{Attempt to push to the ORAM tree}} 
    \State \texttt{NodeMetadata[NodeID].reset()}

\EndProcedure

\end{algorithmic}
\end{algorithm}

Many ORAM protocols have been proposed to protect users' memory traces~\cite{chung2014statistically, gentry2013optimizing, kushilevitz2012security, pathoram, ringoram, hoang2017s3oram}.
This work focuses on an optimized protocol RingORAM~\cite{ringoram}.
Algorithm~\ref{alg:form_pseudocode} shows the pseudocode of the RingORAM protocol. 
RingORAM manages the untrusted cloud memory as a binary tree---ORAM tree. 
Each node in the tree is a bucket and consists of multiple blocks of data at the cache line granularity. 
The blocks are randomly permuted and store up to \texttt{Z} blocks that contain real blocks, and at least \texttt{S} blocks that contain dummy blocks.
All data is encrypted with different keys.
Each physical address \texttt{PA} in the secure logical memory space does not directly correspond to \texttt{PA} on the cloud memory; instead, it maps to a specific \texttt{leaf} (top node) of the ORAM tree.
In the secure domain, the \texttt{PosMap} data structure keeps track of \texttt{leaf} mappings and the actual node location of every \texttt{PA}, and \texttt{Stash} is a small buffer (of typical size 256) that holds the blocks streaming from cloud memory to the trusted processor. 
RingORAM maintains an invariant that \texttt{PA} in the logical memory space lies along the path from its mapped \texttt{leaf} connecting to the root \textbf{or} lies in the \texttt{Stash}.

\begin{figure}[t]
	\centering
		\includegraphics[width=0.48\textwidth, trim={0cm 0cm 0cm 0cm}, clip]{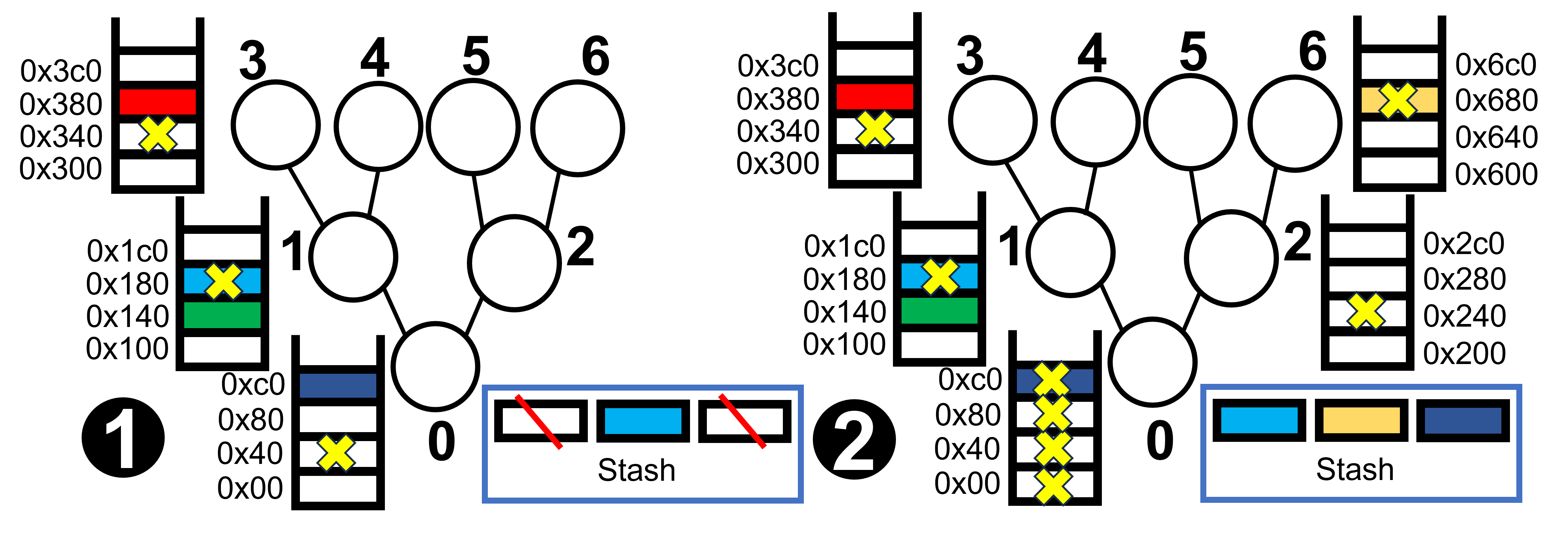}
		\caption{\textbf{\textcolor{black}{
  A toy ORAM access example for illustration purposes. The shown ORAM tree has~\texttt{Z} and~\texttt{S} set to 2. In practice,~\texttt{Z} and~\texttt{S} are much higher. On LLC miss on light blue block, the missed physical address is converted to leaf number to launch accesses along the path and pull blocks into the stash. Once any node is touched~\texttt{S} times, a reset routine is launched. 
        }}
        }
		\label{figure:oram_basics}
\end{figure}

When LLC encounters an R/W miss on a physical address (\texttt{PA}), it queries \texttt{PosMap} to locate the mapped leaf and the position of the node. RingORAM then loads one block from each node along the path from the mapped leaf to the root, generating a stream of memory addresses. Each node selects the real block if it corresponds to the actual location of \texttt{PA}, otherwise a dummy block is chosen. All touched blocks are invalidated for further use~\textcolor{black}{(\circled{\textcolor{white}{1}})}. 
~\textcolor{black}{Fig.~\ref{figure:oram_basics} left shows the example of LLC missing on the light blue block. 
After consulting the \texttt{PosMap} and finding its mapped leaf, exactly one block from each node is streamed into \texttt{Stash} and it is guaranteed to have the block of interest and the rest being dummy blocks. 
}
After loading along the path, \texttt{Stash} decrypts the block of interest to serve the LLC miss, holds the block temporarily, and discards the dummy blocks.
After each access of \texttt{PA}, its mapped leaf is randomly selected again. If \texttt{S} blocks in a node are marked as invalid, violating the future read routine due to the absence of available dummy blocks, the node initiates a reset routine. 
RingORAM first loads \texttt{Z} blocks from the node to \texttt{Stash} and clears the node, then exhaustively iterates through the blocks in \texttt{Stash} to push them back to the reset node~\textcolor{black}{(\circled{\textcolor{white}{2}})}. 
~\textcolor{black}{Fig.~\ref{figure:oram_basics} right shows the example of LLC missing on the yellow block. 
After streaming from the mapped leaf, node 0 is accessed \texttt{S} times and thus requires a reset routine. 
To reset, all blocks in node 0 are streamed into \texttt{Stash} before \texttt{Stash} pushes its content back to the reset node as much as possible. 
All block permutations and valid bits are reset afterward. 
}
Periodic eviction occurs after every \texttt{A} ORAM requests, where a leaf is selected, initiating a reset routine for all nodes along the path from the leaf to the root. RingORAM provides theoretical and empirical evidence that the probability of \texttt{Stash} size overflowing 256 is negligible ($< 2^{-103}$)~\cite{ringoram}.

\subsection{Practical ORAM Implementation}
\label{back:deploy}
\begin{figure}[t]
	\centering
		\includegraphics[width=0.48\textwidth, trim={0cm 0cm 0cm 0cm}, clip]{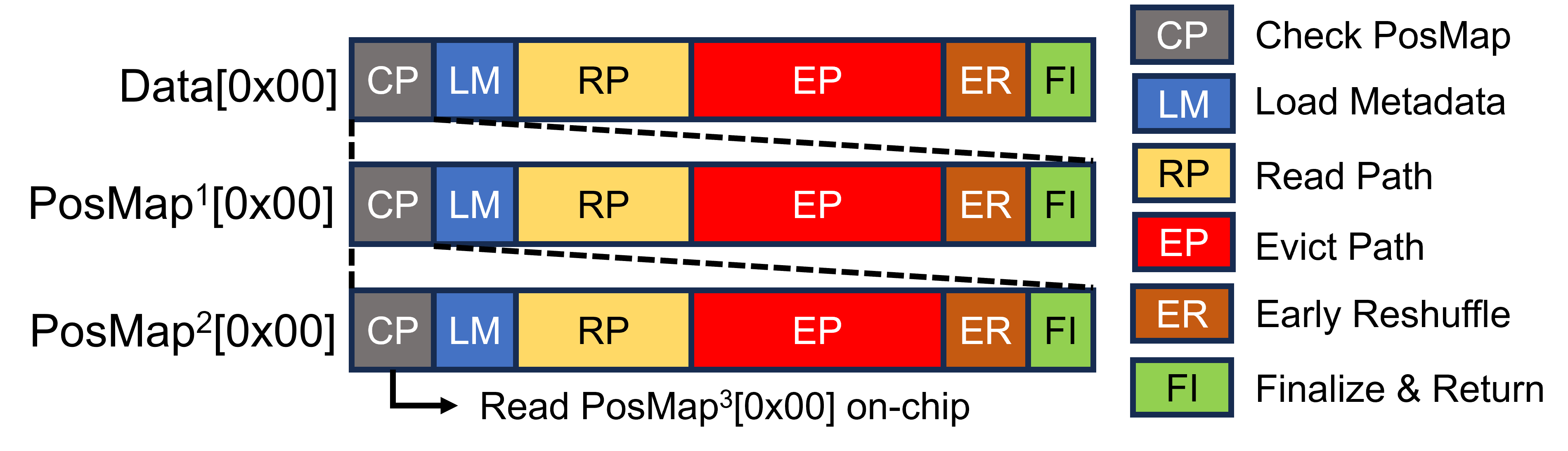}
		\caption{\textbf{Hierarchical ORAM memory spaces. Because the secret data structure \texttt{PosMap} exceeds the on-chip memory capacity, a second-level ORAM protocol is launched to protect the access to data structure \texttt{PosMap}.
        The recursive process continues until \texttt{PosMap} of the protected data structure can be stored on-chip.}
        }
		\label{figure:oram_recursive}
\end{figure}
While protecting a large memory space with ORAM, the capacity of \texttt{PosMap} typically exceeds the on-chip memory capacity on a CPU.
For example, to protect 16GB user space, \texttt{PosMap} takes 2GB of capacity to store the leaf mapping and node position of all data blocks. 
Hence, hierarchical ORAM memory spaces are established to access the encrypted \texttt{PosMap} in the same oblivious access fashion~\cite{fletcher2015freecursive}. 
In Fig.~\ref{figure:oram_recursive}, an example is illustrated, wherein an LLC miss occurs for a 0x00 read operation in the protected user space. 
In response, the ORAM controller initiates a \texttt{PosMap} check for block 0x00.
This launches a sub-ORAM read of the data structure \texttt{PosMap} at 0x00. 
The sub-ORAM launches recursively until the recursive \texttt{PosMap} can be stored on-chip.
Only after the \texttt{PosMap} access, the protocol can find the mapped leaf and resume the RingORAM algorithm execution.

To distinguish the recursive \texttt{PosMap}s, we name \texttt{PosMap$^1$} that keeps track of the protected memory space, and \texttt{PosMap$^2$} that keeps track of \texttt{PosMap$^1$}, and so on. 
Similar to prior works, we use 3 levels of \texttt{PosMaps} and store \texttt{PosMap$^3$} on-chip, the same as all prior works~\cite{pathoram, ringoram, iroram}.
RingORAM imposes a dependency between levels of \texttt{PosMap}s and distinct requests.
In other words,  when presented with requests R1 and R2 for the protected memory space, RingORAM accesses recursive sub-ORAMs to fulfill R1 (\texttt{PosMap$^2$}, \texttt{PosMap$^1$}, \texttt{Data}) first and then proceeds to R2 (\texttt{PosMap$^2$}, \texttt{PosMap$^1$}, \texttt{Data}) in sequence.

\section{Analysis of Prior ORAM Proposals} \label{section:motivation}
\subsection{Analysis of Classical ORAM Implementations}
\label{motivation:perf}

Considering the memory-intensive nature of ORAM, it might be assumed that it fully utilizes the available memory bandwidth.
Moreover, since RingORAM reduces the number of memory accesses, one might intuitively expect it to proportionally outperform the PathORAM baseline.
\textit{However, we discover that this is not the case.}
The RingORAM protocol only marginally outperforms PathORAM by 10\% despite a significant 42\%  of reduction in the number of accesses.

\begin{figure}[t]
	\centering
		\includegraphics[width=0.48\textwidth, trim={0cm 0cm 0cm 0cm}, clip]{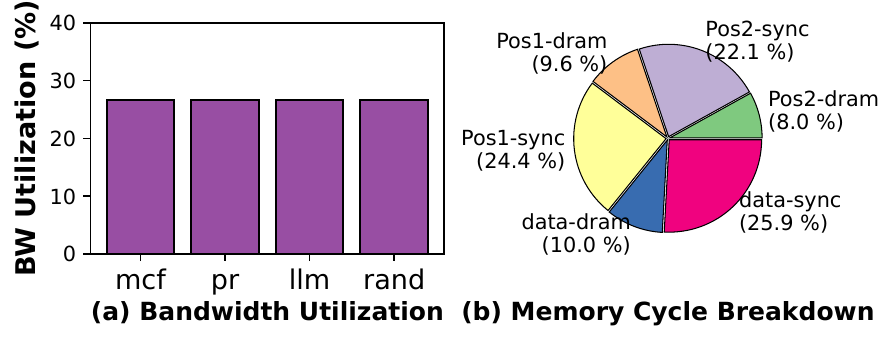}
		\caption{
\textbf{RingORAM protocol bandwidth utilization and performance breakdown. 
        \textcolor{black}{RingORAM incurs less than 30\% bandwidth utilization, which is similar across workloads due to the application of the ORAM protocol.}
        ORAM-sync overhead accounts for 72.4\% of the execution time, which indicates the memory stays idle and spends most of the time waiting for long-latency pull requests to be serviced.}
        }
		\label{figure:Ringoram_breakdown}
\end{figure}

Fig.~\ref{figure:Ringoram_breakdown} shows the performance breakdown of RingORAM using detailed methodology discussed in \S\ref{section:methodology}.
\textcolor{black}{The baseline ORAM controller issues all resulting reads and incurred stash evictions of an ORAM request to the memory controller. 
Without waiting for the writes to commit, the ORAM controller keeps issuing the next ORAM request to saturate the memory controller queues for memory-level parallelism. 
In the RingORAM protocol, ORAM requests are served one after another to avoid a block being read by multiple requests before being updated. 
Otherwise, it gives the attacker accurate information about the dummy location and violates obliviousness}. 
For a single request, after selecting the leaf-to-root path, the request needs to load metadata on-chip to determine which blocks in the nodes are usable.
The data load can only start after their dependent metadata are loaded. 
Additionally, the protocol specifies that \texttt{Stash} can only push blocks to the ORAM tree after \texttt{ReadPath} and \texttt{ResetBucket} pull all their required blocks. 
These requirements introduce substantial synchronization overhead, as pull requests are held up while waiting for their dependent metadata to be in place, and the \texttt{Stash} is delayed until all pull requests are in position.

We refer to the memory controller stall cycles resulting from these reasons as ORAM-sync cycles.
Surprisingly, ORAM-sync accounts for around 72.4\% of the execution time.
In essence, RingORAM dedicates a substantial portion of time to stalling the memory controller, awaiting the completion of preceding high-latency pull requests before it can advance with the protocol.
\textcolor{black}{These stalls are inherent to executing the RingORAM protocol, even with a multi-issue ORAM controller attempting to saturate the memory bandwidth.
Notably, the average DRAM bandwidth utilization remains below 30\% across all workloads. 
\textcolor{black}{
We additionally provide the approximate analytical calculation to support the cycle-accurate simulation results. 
The DRAM request latency for row-hits and row-misses are tCL and (tCL+tRP+tRCD), respectively. We find that 48.2\% of the requests are row-hits in RingORAM. Furthermore, the average memory controller queue occupancy across all channels is 21.1 due to frequent dependency stalls of the protocol. Using DRAM timing parameters for DDR4-3200, the average bandwidth we find is 64B $\times$ 21.1 / 46.9ns = 28.8GB/s (theoretical-maximum: 102.4GB/s): close to 28.1\% bandwidth utilization. 
}
Note that despite the varying nature of these workloads in terms of memory traffic, each LLC miss request undergoes conversion into a single ORAM request by mapping to a unique ORAM tree leaf on every occasion.
The homogenization of memory bandwidth utilization occurs due to applying the ORAM protocol, resulting in indistinguishable memory traffic across different workloads, thereby achieving the primary objective of ORAM.
}

The access to sub-ORAMs accounts for 64.1\% of the time, which is due to the capacity limitation of on-chip structures (\S\ref{back:deploy}).
Additionally, the protocol forces these accesses to issue and complete in order, servicing one request at a time.
This study reveals that there is an untapped potential for memory-level parallelism in serving different requests because different \texttt{PosMap}s are protecting different memory spaces.
For instance, concurrently serving accesses R1 \texttt{PosMap$^1$} and R2 \texttt{PosMap$^2$} does not result in conflicts. 
No prior work takes advantage of this optimization opportunity.

\textcolor{black}{
\subsection{Prefetch-based Optimization Strategies}
\label{section:prefetch_opt}
To improve ORAM performance, many prior works~\cite{iroram, proram, rajat2023laoram} propose to eliminate ORAM requests. 
For example, PrORAM and its variants~\cite{che2020multi, proram, rajat2023laoram} use PathORAM~\cite{pathoram}, a prior state-of-the-art protocol, as their baseline. 
PathORAM manages the ORAM tree similarly to RingORAM, with a key distinction being that in PathORAM, a \texttt{node} does not differentiate between real or dummy blocks. 
Each LLC miss on \texttt{PA} in PathORAM results in loading \textbf{all} blocks along the \texttt{nodes} connecting from its mapped \texttt{leaf} to the root.
Additionally, upon loading from a \texttt{leaf}, PathORAM evicts the same \texttt{leaf} path immediately. 
The PrORAM protocol~\cite{proram} forces the mapping of consecutive physical addresses in the original memory space to the same leaf in the ORAM tree. 
This way, one load of a physical address effectively \textit{prefetches} multiple data blocks from the ORAM tree to \texttt{Stash} and then to LLC. 
Subsequent accesses, if hit the LLC, bypass the ORAM protocol.
This solution achieves a notable speedup by mitigating the ORAM workload when the original memory access exhibits high locality.
}

\textcolor{black}{However, PrORAM optimization suffers from two limitations, which we discuss through qualitative and quantitative analyses as follows. 
First, the prefetch performance benefit is sensitive to the locality pattern. 
This optimization has little effect on workloads with low to moderate locality in their original LLC miss trace such as some SPEC benchmarks, real-world graph analytics, and random key-value accesses, with demonstrated in performance evaluation in \S\ref{perf:main}. 
Second, forcing the mapping of consecutive physical addresses in the original memory space to the same leaf in the ORAM tree limits the scope that \texttt{Stash} can distribute its blocks back to the ORAM tree. 
This violates the premise in PathORAM proof that blocks are assigned ``independently and uniformly at random" into leaves of the ORAM binary tree~\cite{pathoram}. 
As a result, background eviction is introduced when \texttt{Stash} overflows in this scenario~\cite{ren2013design}.}

\begin{figure}[t]
	\centering
		\includegraphics[width=0.48\textwidth, trim={0cm 0cm 0cm 0cm}, clip]{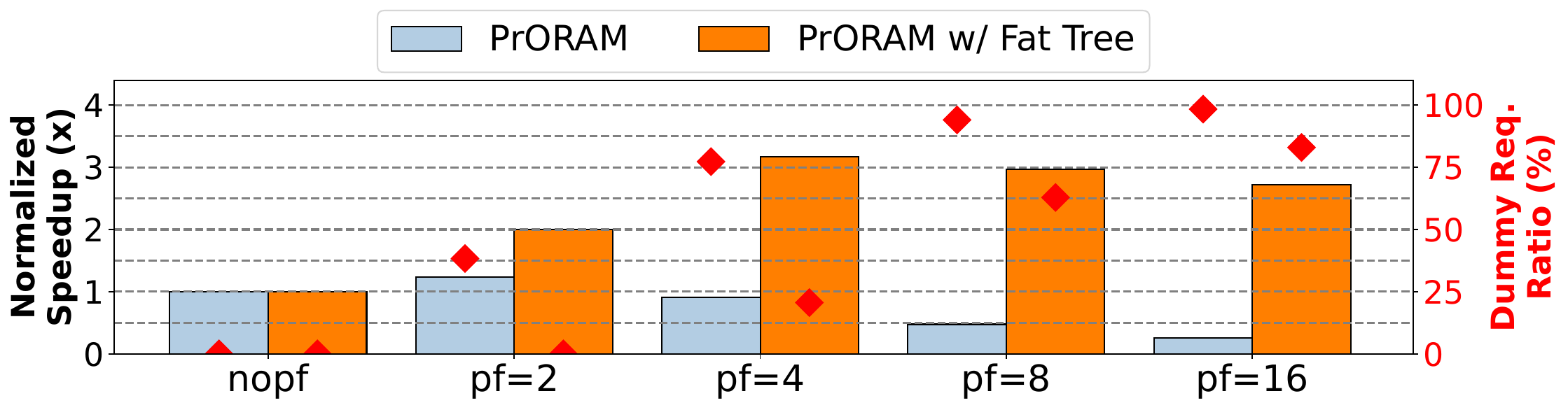}
		\caption{
        \textbf{\textcolor{black}{Normalized speedup of PrORAM and LAORAM (PrORAM w/ Fat Tree) running \streams, a synthetic workload where consecutive cache line addresses are missed subsequently by the LLC. pf=X refers to forcing mapping to the same leaf for a prefetch length of X. A high dummy request ratio limits the performance scaling despite the present locality. }}
        }
		\label{figure:pr_movitaion}
\end{figure}
\textcolor{black}{Whenever the \texttt{Stash} size exceeds the threshold, a dummy ORAM request is inserted to read from and write to a random dummy path. 
This path access does not contribute to fulfilling any LLC requests but only helps to further clear the \texttt{Stash}. 
Fig.~\ref{figure:pr_movitaion} shows the quantitative results of how these dummy requests affect performance. 
The experiment models PrORAM with a 1024-entry \texttt{Stash} protecting \streams, a synthetic workload where consecutive cache line addresses are missed subsequently by the LLC. 
Ideally, a higher prefetch length configured in PrORAM should perform better than a lower one because the perfect locality of the trace can eliminate more of the ORAM requests. 
However, we find that PrORAM performance does not scale with the prefetch length.}

\textcolor{black}{At a prefetch length of 4, although the original ORAM leaf accesses are reduced by 4$\times$, the \texttt{Stash} pressure has to repeatedly insert dummy requests, and the fraction of dummy requests accounts for 77.3\% of the total ORAM requests. 
This leads to even a slowdown compared to the no prefetch case. 
PrORAM proposes to use a threshold to adapt to the background eviction frequency and dynamically disable or limit the prefetch length. 
This achieves up to 1.5$\times$ speedup over the no prefetch case (see Fig. 7 in PrORAM~\cite{proram}).}

\textcolor{black}{LAORAM proposes a Fat-Tree structure to allocate 2$\times$ block size at the root level and gradually decrease the bucket size going up towards the leaf~\cite{rajat2023laoram}. 
This significantly reduces the \texttt{Stash} pressure and decreases the dummy request ratio. 
However, its performance is capped at 3.2$\times$ at a prefetch length of 4 even with perfect locality in \streams.
Note that LAORAM reports a few over 3.2$\times$ speedups in some workloads. 
This is because LAORAM uses a software-managed \texttt{Stash} and allows the \texttt{Stash} to grow above 3600 after 12500 ORAM accesses and remain unbounded throughout the workload lifetime (see Fig. 8 in LAORAM~\cite{rajat2023laoram}). 
This reduces the dummy request ratio observed in LAORAM. 
However, because \textit{all} entries in \texttt{Stash} need to be probed on every ORAM access, a high-performance on-chip ORAM solution requires \texttt{Stash} to be in hardware and remain small due to the need for area efficiency and probing with high associativity~\cite{maas2013phantom}. 
Such an unbounded \texttt{Stash} cannot be justified in a high-performance on-chip hardware solution. 
}

\subsection{Summary of Challenges Optimizing ORAM Performance}
The ORAM performance is limited due to its memory-intensive nature, where a single memory request in the unsecured domain is converted into 100s of memory requests~\cite{pageoram,ringoram}.
As a response, prior works~\cite{iroram, proram, rajat2023laoram,che2020multi} attempt to improve performance by bypassing the ORAM protocol for a subset of memory requests.
\textcolor{black}{While these designs yield some performance improvement, the benefits are highly sensitive to workload behavior.
Additionally, we uncover that the benefit of these optimizations does not scale with prefetch length even with perfect locality due to frequent \texttt{Stash} overflows and the introduction of dummy ORAM requests.
The primary challenge in enhancing the performance of ORAM lies in accelerating a broad class of workloads with varying locality patterns and avoiding dummy requests.
}
\THISWORK\ addresses this open research question.


\section{Introducing Concurrency using \THISWORK}
\label{section:hwsfinterface}



\subsection{Key Design Goals}
The primary goal of \THISWORK\ is to improve performance by enhancing the memory bandwidth utilization without sacrificing security.
The goals of \THISWORK\ design are:

\begin{itemize}[leftmargin=*]
    \item \textit{Simultaneously processing multiple ORAM requests.}
    As discussed in \S\ref{motivation:perf}, accessing multiple ORAM requests concurrently is the key to overcoming the long ORAM request latency and improving bandwidth utilization.
    \item \textit{Minimizing execution bubbles.}
    Facilitating concurrent access to multiple ORAM requests necessitates identifying minimal dependencies between them, as this is integral to achieving high-performance ORAM while preserving memory functionality correctness and obliviousness.   
    \item \textcolor{black}{\textit{No compromise on security.}
    The protocol must be carefully designed such that any concurrent access behavior does not \textcolor{black}{compromise the security guarantee of the original protocol}.}
\end{itemize}

\subsection{Unlocking Parallelism in ORAM Protocol}
\label{section:oram_parallelism}
\begin{figure}[t]
	\centering
		\includegraphics[width=0.48\textwidth, trim={0cm 0cm 0cm 0cm}, clip]{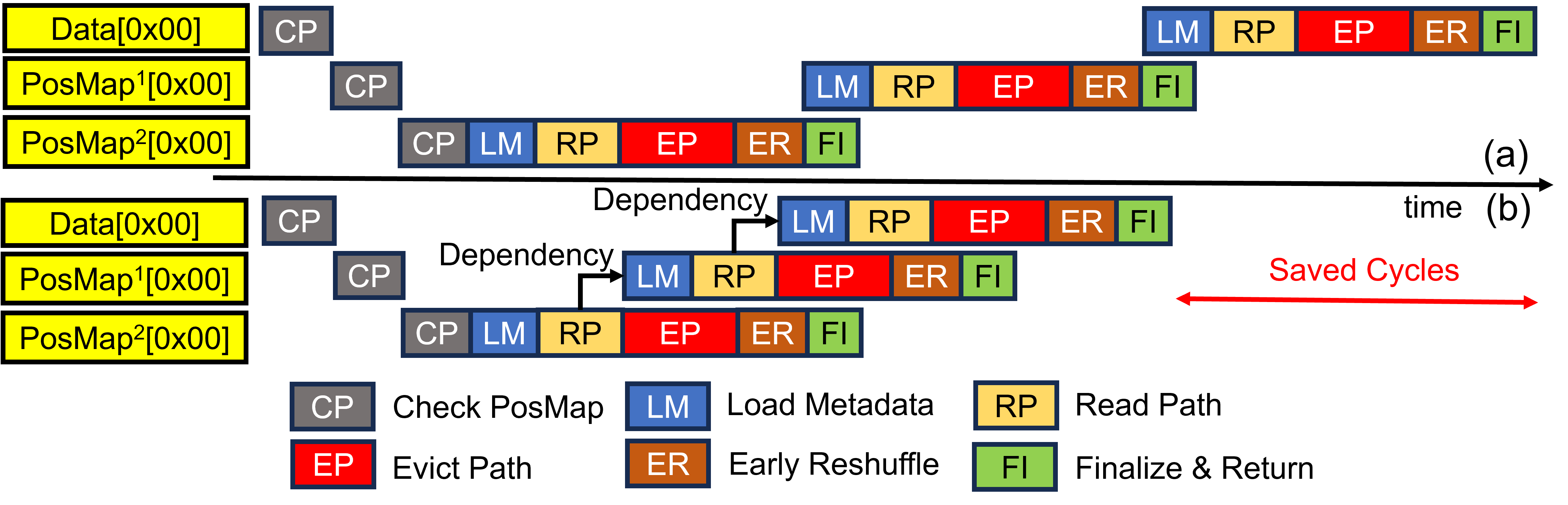}
		\caption{
        \textbf{Intra-request parallelism in serving a single ORAM request.}
        }
		\label{figure:intra_parallel}
\end{figure}

\begin{figure}[t]
	\centering
		\includegraphics[width=0.48\textwidth, trim={0cm 0cm 0cm 0cm}, clip]{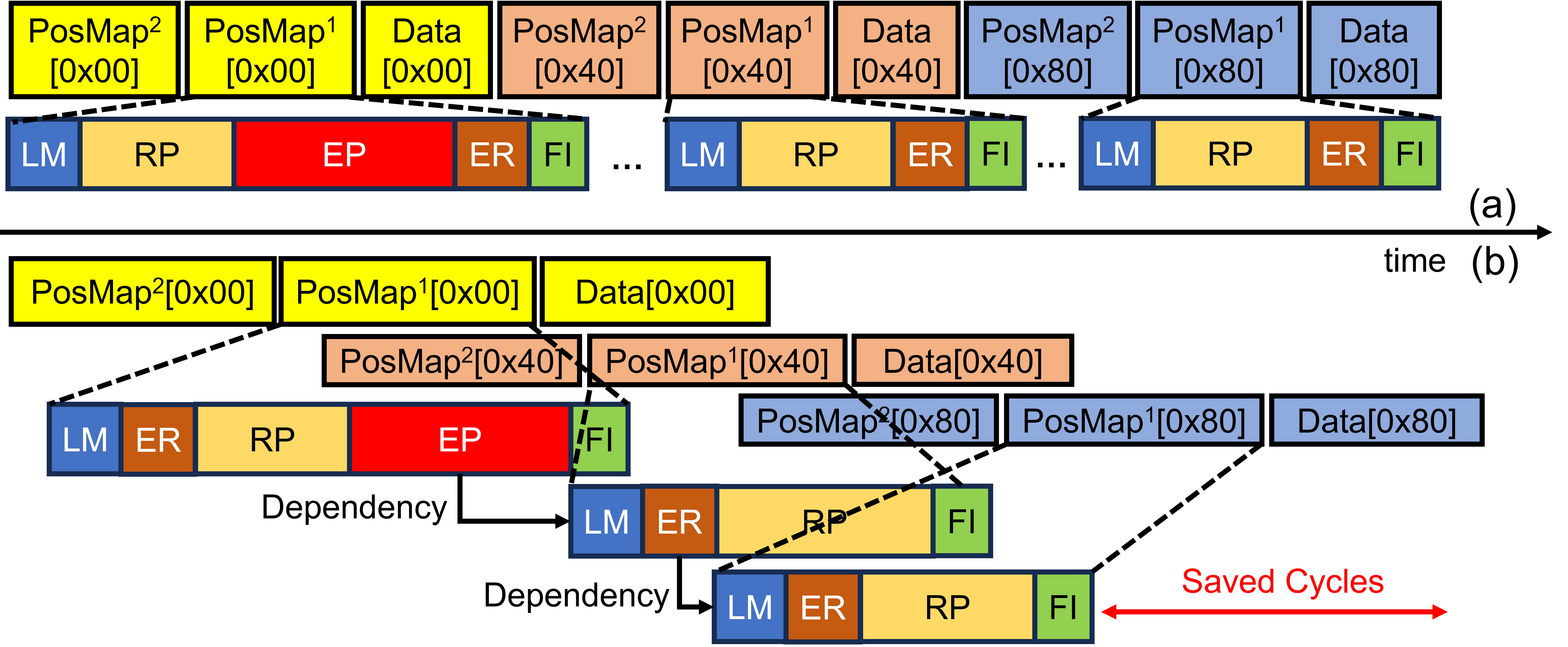}
		\caption{
        \textbf{Inter-request parallelism in serving multiple ORAM requests.}
        }
		\label{figure:inter_parallel}
\end{figure}

To enhance performance, we analyze the minimal dependencies between ORAM requests that the ORAM controller must adhere to for correctness. 
We crucially observe that there are two categories of achievable parallelism to concurrently serve ORAM requests while ensuring correct functionality.

\textbf{Intra-Request Parallelism.}
In Fig.~\ref{figure:intra_parallel}(a), the baseline RingORAM protocol serving a single ORAM request of 0x00 with a hierarchical design is illustrated.
Each level of ORAM launches a recursive sub-ORAM to check the PosMap value of the requested block. 
We make a crucial observation that each sub-ORAM memory space is exclusive; in other words, the ORAM memory space to protect \texttt{PosMap$^2$} is entirely exclusive from \texttt{PosMap$^1$}.
Thus, accesses to one level of sub-ORAM can be concurrently executed with accesses to another.
Fig.~\ref{figure:intra_parallel}(b) shows the minimal dependency that needs to be respected when serving different sub-ORAMs for a single request. 
\texttt{PosMap$^1$} \texttt{LM} step can start as soon as the mapped leaf is known. 
This is resolved as soon as \texttt{PosMap$^2$} \texttt{RP} is completed. 
The write to \texttt{PosMap$^2$} ORAM tree (\texttt{EP}) and the read of \texttt{PosMap$^1$} ORAM tree (\texttt{RP}) can be executed in parallel without any conflicts because they access exclusive memory spaces.
Using these observations, we design the \THISWORK\ protocol that unlocks intra-request parallelism.

\textbf{Inter-Request Parallelism.}
Fig.~\ref{figure:inter_parallel}(a) shows the baseline RingORAM protocol serving multiple ORAM requests. 
Assuming LLC misses on addresses 0x00, 0x40, and 0x80,
in the baseline, ORAM access to \texttt{PosMap$^2$}[0x00] executes the RingORAM protocol in order followed by the access to \texttt{PosMap$^1$}[0x00]; ORAM access to \texttt{PosMap$^2$}[0x40] is serialized after the access to \texttt{Data}[0x00]. 
We observe that sub-ORAMs on the same level (\textit{e.g.}, \texttt{PosMap$^1$}[0x00] and \texttt{PosMap$^1$}[0x40]) can be concurrently accessed in most conditions, even though they share the same memory space.
This is a distinctive characteristic of RingORAM because the protocol ensures that it invalidates precisely one data block position in the node upon being touched, and subsequent access to the node always selects a different position in the node if available.
This unique characteristic of the RingORAM protocol unlocks massive opportunities for parallelism. 
This exclusivity access guarantee, however, can be violated once the node resets.
This is because the reset operation pushes values from the stash and alters the content of the ORAM tree, and subsequent requests must have the updated view of the tree to maintain correctness.
Hence, the modification of the ORAM tree (\texttt{EP} and \texttt{ER}) by one ORAM request and the read (\texttt{LM} and \texttt{RP}) by the subsequent request forms the minimal critical section for concurrently accessing multiple requests on the same sub-ORAM level.
The \texttt{ER} step is executed on every access, whereas \texttt{EP} is executed after every \texttt{A} access. 

To overcome this challenge, we propose to re-order and hoist the execution of the \texttt{ER} step to the earliest stage possible.
The purpose of this is to resolve the write-to-read critical section as soon as possible.
This approach allows the current request to modify the ORAM tree at the earliest opportunity to a ``good to read" state and then pass the tree to the subsequent request.
Meanwhile, \texttt{EP} is serialized after \texttt{RP} to uphold the theoretical guarantee that the stash remains bound to a fixed size, regardless of the concurrency order.
In Fig.~\ref{figure:inter_parallel}(b), our proposed protocol is depicted to minimize the critical path of concurrently running multiple requests.
Consider the \texttt{PosMap$^1$} ORAM tree, the execution of \texttt{PosMap$^1$}[0x40] can be issued as soon as \texttt{PosMap$^1$}[0x00] \texttt{EP} stage is complete. 
The execution of \texttt{PosMap$^1$}[0x80] can be issued once \texttt{PosMap$^1$}[0x40] \texttt{ER} stage is complete.
All \texttt{RP} (non-store) steps from different requests in the same memory space can also be overlapped, which saves a significant amount of cycles.
This design eliminates unnecessary dependencies.

\subsection{\THISWORK\ Protocol Details}

\begin{algorithm}[t]
\scriptsize
\caption{Pseudocode for \THISWORK\ algorithm to serve ORAM requests concurrently. Changes are marked in red. }
\label{alg:software_pseudocode}
\begin{algorithmic}[1]
\Procedure{\textcolor{blue}{\THISWORK\ ORAMAccess}}{\texttt{PA, op, data', GlobalID}} \\
    \hspace*{\algorithmicindent} \textbf{Input}: \texttt{GlobalID}: Global issue ID when accessing concurrently \\
    \hspace*{\algorithmicindent} \textbf{Global}: \texttt{CommitHead}: Synchronization for original memory request order   

    \State \texttt{\textcolor{red}{while(CommitHead != GlobalID) \{;\}}} \Comment{\textcolor{red}{sleep and wait for sync}}
    \State \texttt{leaf = UniRandomLeaf if PA \textcolor{red}{pending} else PosMap[PA]}
    \State \texttt{leaf' = UniRandomLeaf}
    \State \texttt{\textcolor{red}{Mark PA as pending}}
    \State \texttt{PosMap[PA] = leaf'}

    \State \texttt{\textcolor{red}{{EarlyReshufflePreCheck(leaf)}}}
    \If{\texttt{GlobalID \% A != 0}}
    \State \texttt{\textcolor{red}{AtomicAdd(\&CommitHead, 1)}}
    \EndIf
    \State \texttt{data = ReadPath(leaf, PA)} 
    
    
    \If{\texttt{op == READ}}
        \State \texttt{return data to processor from Stash}
    \Else
        \State \texttt{Stash[PA] = data'}
    \EndIf
    
    \If{\texttt{GlobalID \% A == 0}}
        \State \texttt{EvictPath()}
        \Comment{\textcolor{red}{Remove pending status upon blocks evicted from stash}}
        \State \texttt{\textcolor{red}{AtomicAdd(\&CommitHead, 1)}}
    \EndIf

    \State \texttt{\textbf{return}}
\EndProcedure
\\
\Procedure{\textcolor{blue}{EarlyReshufflePreCheck}}{\texttt{leaf}} 
    
    \ForAll{\texttt{NodeID} $\in$ \texttt{leaf along root}}
    \If{\texttt{NodeMetadata[NodeID].accessed == \textcolor{red}{S - 1}}}
    \State \texttt{ResetBucket(NodeID)}
    \Comment{\textcolor{red}{Mark Node as bypassed in ReadPath()}}
    \EndIf
    \EndFor
\EndProcedure
\end{algorithmic}
\end{algorithm}

Algorithm~\ref{alg:software_pseudocode} presents pseudocode for the proposed \THISWORK\ protocol that includes the following changes.
First, we elevate the execution order of \texttt{EarlyReshuffle} to preserve correctness.
Second, \THISWORK\ protocol serves an arbitrary number of requests concurrently while using \texttt{CommitHead} to synchronize the memory serving order to the same order they are issued.
This ensures that using concurrent ORAM does not introduce any memory consistency issues.
For each request, the \texttt{CommitHead} waits for all previous tree locks to be released.
Then, it locks the ORAM tree until all operations that possibly modify the ORAM tree finish.

\THISWORK\ protocol (\THISWORK\--SW) enables concurrent ORAM accesses and exploits the inter-request parallelism as multiple \texttt{ReadPath()} can be overlapped.
However, the coarse-grained nature of software synchronization limits the benefit of \THISWORK.
For example, the synchronization primitive placed around the \texttt{PosMap} check prevents the intra-request parallelism.
To address this, we present the design of ORAM controller architecture to exploit parallelism at both levels.

\section{\THISWORK\ ORAM Controller Design} \label{section:hardware_design}

This section presents the hardware design of an ORAM controller to fully support the memory-level parallelism introduced by \THISWORK\ protocol.

\begin{figure}
    \centering
    \includegraphics[width=0.48\textwidth]{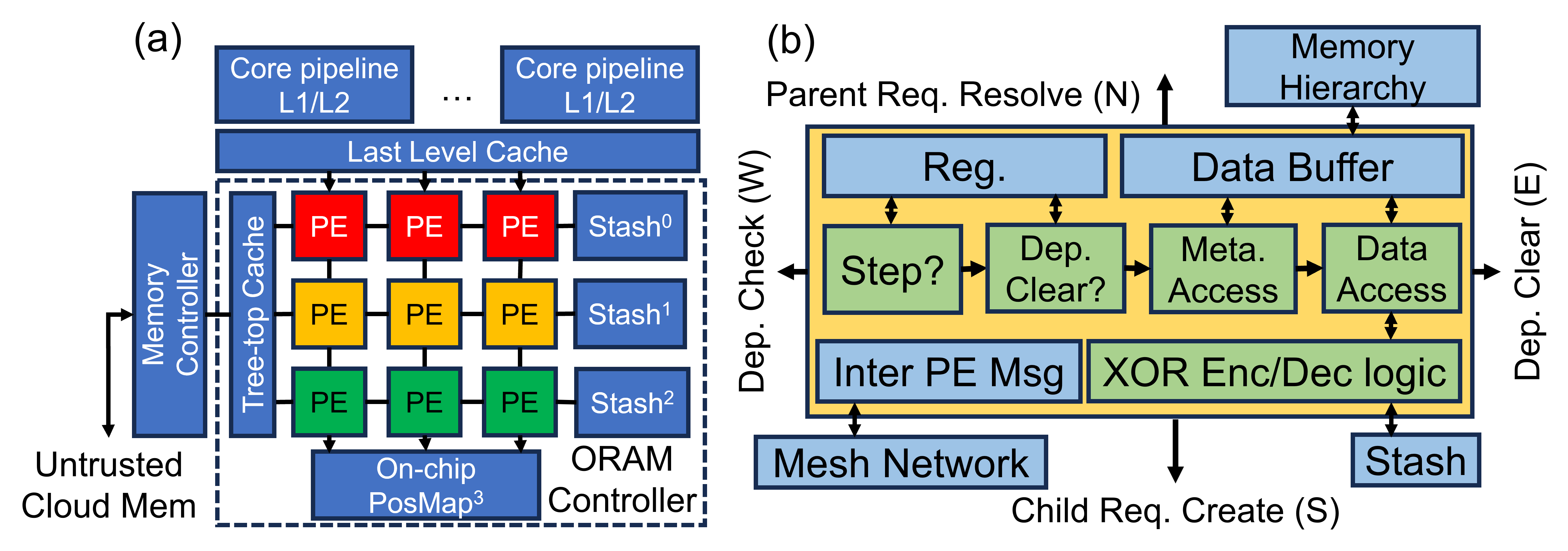}
    \caption{\textbf{(a) \THISWORK\ ORAM controller consists of a 2D array of PEs, where each row of PE serves separate hierarchical sub-ORAMs, and each column of PE serves multiple ORAM requests concurrently, and (b) a single PE architecture for \THISWORK.}}
    \label{fig:pe}
\end{figure}

\subsection{Hardware Support for Unlocking Additional Parallelism}
Fig.~\ref{fig:pe} shows the proposed hardware architecture of \THISWORK\ ORAM controller. 
The architecture consists of 3 levels of Processing Elements (PEs).
Each row of PEs serves a separate protected memory space of \texttt{Data}, \texttt{PosMap$^1$}, and \texttt{PosMap$^2$}.
Each column of PEs serves a single request issued from an LLC miss.
For example, three misses will be mapped to three columns of PEs.
Within each PE, the finite state machine executes Algorithm~\ref{alg:software_pseudocode} independently.
Protocol-level dependencies are managed by the communication with neighboring PEs.
The execution flow operates as follows.

\noindent
\textbf{Check \texttt{PosMap}:} PE receives the query for \texttt{PosMap} from the parent (from the north) and sends it to the child (to the south), awaiting the response from the child to retrieve the leaf ID mapped to the requested data block.

\noindent
\textbf{Load Metadata:} PE waits until the sibling dependency from the west is cleared. 
PE then loads metadata from the leaf to the root of the ORAM tree.

\noindent
\textbf{Early Reshuffle:} 
Checking the loaded metadata, PE resets all nodes along the path that need an early reshuffle. 
After all reshuffles are issued, it clears the dependency for the sibling to the east (\textit{i.e.,} ORAM tree is ready for next read).

\noindent
\textbf{Read Path:}
Processing the metadata, PE issues requests and loads data along the leaf to root, decrypts blocks, and responds to the parent with the values of the requested block. 

\noindent
\textbf{Evict Path:} 
If there is an eviction being scheduled, the stash is re-encrypted and an eviction process is initiated. 
This clears the dependency for the sibling to the east. 

\noindent
\textbf{Finalize:} PE is ready to retire and clears up when all rows (sub-ORAMs) of the same request are finalized. 

All PEs execute an identical workflow and concurrently issue memory requests to enhance bandwidth utilization. 
~\textcolor{black}{The PE array forms a ring, and once a new request occupies the left-most PE, the right-most PE sends clear signals to the left-most PE regarding sibling dependency. }

\subsection{A Walk-Through Example}
\begin{figure}
    \centering
    \includegraphics[width=0.4\textwidth]{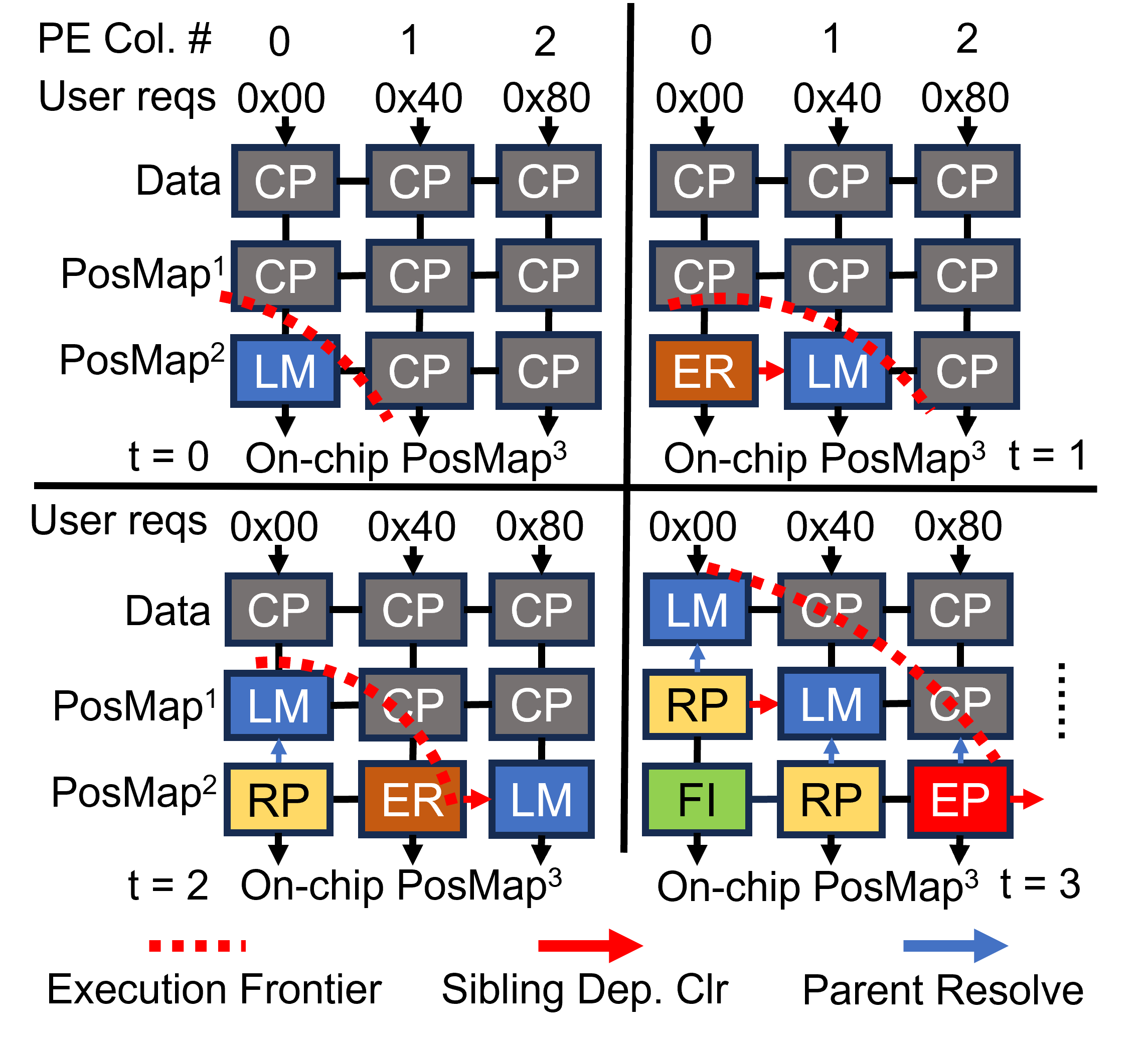}
    \caption{\textbf{\THISWORK\ architecture walkthrough example.}}
    \label{fig:2d_pipeline}
\end{figure}

Figure~\ref{fig:2d_pipeline} illustrates the proposed hardware's operation with a 3$\times$3 PE mesh. 
Initially, at t = 0, LLC has three outstanding misses: 0x00R, 0x40R, and 0x80R.
\THISWORK\ controller first registers all 3 requests in a separate column in \texttt{Data} PE.
The on-chip storage of \texttt{PosMap$^1$} for address 0x00R is absent, resulting in the designation of the \texttt{Data}-0 PE phase as \texttt{CP}, initiating an access to address \texttt{PosMap$^1$}[0x00] in the sub-ORAM.
This launches another recursive access, setting the \texttt{PosMap$^1$}-0 PE the \texttt{CP} phase.
This process continues until the position map is found on-chip at the third level.
\texttt{PosMap$^2$}-0 PE accesses this on-chip mapping and progresses to the \texttt{LM} phase.
While this request progresses, other requests are still stalled at the \texttt{CP} phase.
At t = 1, \texttt{PosMap$^2$}-0 PE finishes the \texttt{ER} phase.
This signals the completion of the writing phase of \texttt{PosMap$^2$}, prompting the transmission of a dependency clear signal (red arrow) to the PE to the east, instructing \texttt{PosMap$^2$}-1 to execute \texttt{LM}.
As the \THISWORK\ protocol ensures that the subsequent execution of \texttt{PosMap$^2$}-0 and \texttt{PosMap$^2$}-1 will not access the same address, these two PEs can operate concurrently.
At t = 2, \texttt{PosMap$^2$}-0 completes \texttt{RP}, transmitting the \texttt{CP} request response to the requester \texttt{PosMap$^1$}-0, while \texttt{PosMap$^2$}-1 finishes \texttt{ER}, unlocking \texttt{PosMap$^2$}-2 to execute \texttt{LM}.
Note that for PE of request-id being multiples of \texttt{A}, the ORAM tree write phase is complete only when \texttt{EP} is complete.

The execution frontier moves in a waveform and unlocks a massive level of parallelism across different requests with minimal dependencies. 
At the column level, multiple requests can issue the time-consuming stage \texttt{ReadPath()} concurrently. 
Upon issuing the ORAM tree modification of the current request to the memory controller, the neighboring PE in the east can promptly start issuing DRAM requests without waiting for the modification to complete.
This uplifts the memory bandwidth utilization rate significantly.
At the row level, different sub-ORAMs can issue requests at the same time.
When the ORAM tree has completed the \texttt{ReadPath()} phase, the parent request can immediately start to issue.

\subsection{\textcolor{black}{\THISWORK\ Integration with Prefetching}} \label{subsection:palermo_prefetch}
\textcolor{black}{
Inspired by prior works~\cite{che2020multi, proram, rajat2023laoram}, \THISWORK\ can also support prefetching to further improve performance.
To enable prefetching of contiguous data blocks, \THISWORK\ maps multiple data blocks/cache lines to a single \texttt{Data} block in the ORAM tree.
This is achieved seamlessly upon address translation from \texttt{Data} ORAM tree block accesses in \texttt{RP} phase to a sequence of \textit{prefetch length} DRAM accesses, while \texttt{PosMap$^1$} and \texttt{PosMap$^2$} ORAM tree accesses remain unchanged. 
As shown in Fig.~\ref{figure:stash}, \THISWORK\ prefetch scheme does not increase the \texttt{Stash} pressure as opposed to PrORAM and thus eliminates the necessity for injecting any dummy ORAM requests.
Prefetching does not change the \THISWORK\ protocol and only requires a wider \texttt{Stash} design to accommodate wider data blocks in hardware.
Notably, prefetching is not a necessary design choice, but can be optionally used with \THISWORK\ based on ideas presented in prior works~\cite{che2020multi, proram, rajat2023laoram}.
This feature is configured by the user before execution and is fully decoupled with the underlying protected workload behavior. }

\subsection{Discussion: Comparison with \textcolor{black}{Software-based Parallel ORAM Protocols}} \label{subsection:concuroram}

\textcolor{black}{
Software-based ORAM optimizations~\cite{chakraborti2018concuroram, sahin2016taostore, williams2012privatefs} have been proposed to query untrusted databases.
To unlock parallel transaction processing for multiple clients, ConcurORAM issues multiple ORAM requests one after the other with ``order-based synchronization'' without waiting for their evictions to be committed. 
All evictions are tracked in an append-based log and are processed in the background.
}



\textcolor{black}{
Crucially, ConcurORAM is a software library that adopts the ORAM protocol algorithm to access an external untrusted database securely.
While ConcurORAM achieves memory saturation, it refers to achieving peak performance in software. 
In contrast, \THISWORK\ designs a novel ORAM protocol and an underlying hardware implementation to access an untrusted RAM efficiently.
In fact, this co-design unlocks memory-level parallelism at the memory controller to reduce execution bubbles and improve DRAM bandwidth utilization.
}

\subsection{Discussion: \textcolor{black}{Optimizing RingORAM over PathORAM}} \label{subsection:whyringoram}
\textcolor{black}{
RingORAM presents major design improvements over PathORAM to reduce DRAM traffic.
Despite the 42\% less traffic, RingORAM marginally outperforms PathORAM by 10\% mainly due to the blocking issue nature and frequent dependency stalls presented in \S\ref{motivation:perf}.
\THISWORK\ finds the unique opportunity in RingORAM to respect the minimal dependency at a hardware level and exploit higher memory-level parallelism. 
PathORAM, on the other hand, does not have RingORAM's access exclusivity guarantee that subsequent accesses always select distinct blocks in a node if available (\S\ref{section:oram_parallelism}). 
Thus, a similar strategy in PathORAM gains limited performance benefits because PathORAM protocol in its nature has much higher total DRAM traffic with fewer dependency bubbles, leaving little room for improvement at the memory controller.
\S\ref{perf:main} presents the detailed performance analysis of all studied baselines based on the two protocols.
}

\section{Security Analysis of \THISWORK} \label{section:security}


\textcolor{black}{
The following qualitative and quantitative analyses show that
\THISWORK\ design upholds the same security guarantees as the RingORAM protocol.
}

\begin{figure}
    \centering
    \includegraphics[width=0.48\textwidth]{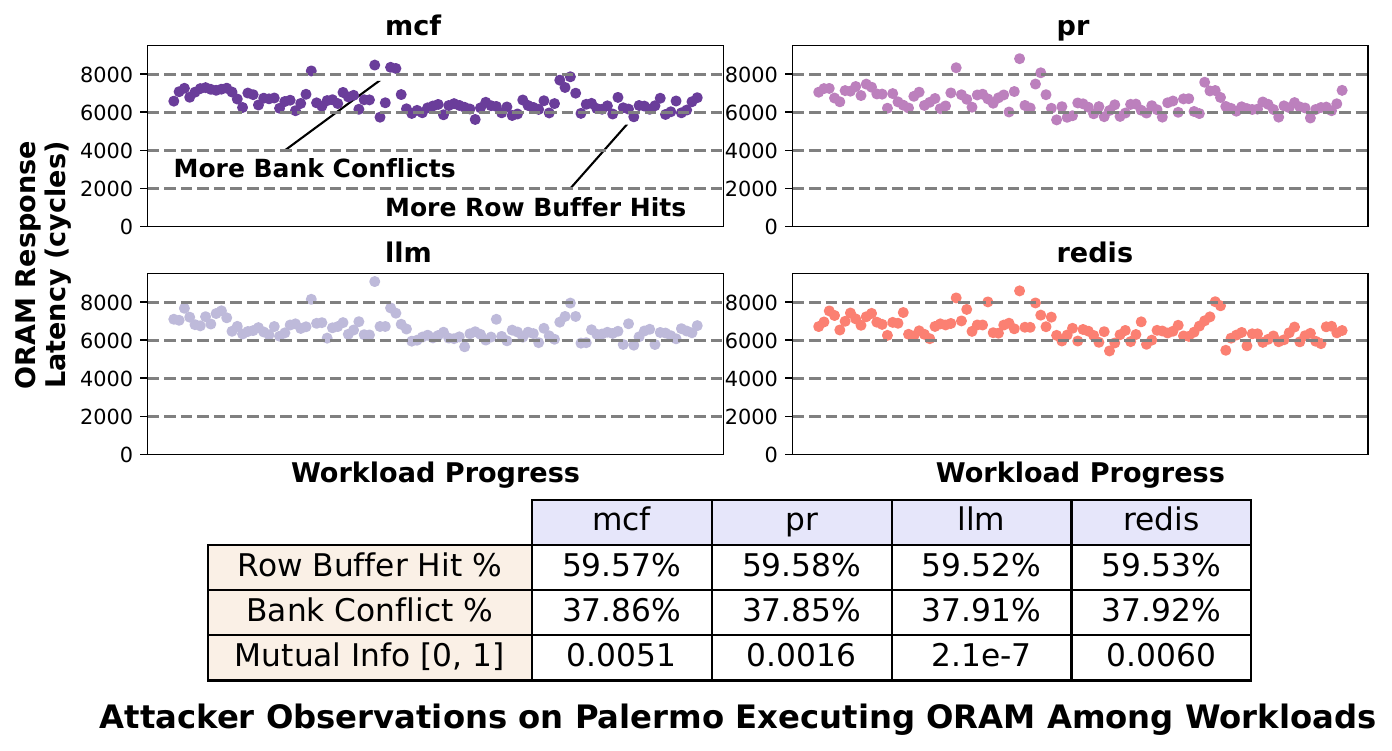}
    \caption{\textbf{\textcolor{black}{ORAM response latency in \THISWORK. The variations are due to the leaf selections, memory controller, and DRAM architecture, all of which are public information. 
    Quantitatively, the mutual information in all workloads is shown in the table. 
    The close to 0 result indicates the attacker's knowledge gain is no better than random in the original program by observing the DRAM timings. 
    Qualitatively, this is expected because all DRAM timings are only determined by the statistically random leaf sequences, thus fully decoupled with any program behavior.
    }}
    }
    \vspace{-15pt}
    \label{fig:security_new_workloads}
\end{figure}

\textbf{Qualitative analysis.}
\textcolor{black}{
We first qualitatively show that \THISWORK's access traces are indistinguishable.
In \THISWORK, the ORAM controller issues LLC misses at a constant rate and pads dummy ORAM requests when LLC issues none, same as~\cite{iroram}. 
Each ORAM request \texttt{PA} in \THISWORK\ protected memory space is mapped to an ORAM leaf.
Upon accessing this leaf as it becomes visible on the DRAM memory bus, \THISWORK\ immediately re-maps the \texttt{PA} to another random ORAM leaf without revealing it, following the remapping scheme from RingORAM.
Thus, an attacker monitoring the memory traffic that executes DRAM bus attacks can only infer the underlying stream of statistically random leaf selections at a constant rate and the statically configured \textit{prefetch length}, both of which are uncorrelated with the original program behavior.
In conclusion, \THISWORK\ follows the same level of security guarantee as RingORAM in hiding the memory access patterns.
}

\textbf{Quantitative analysis.}
\textcolor{black}{
\THISWORK's distinct characteristic from other ORAM optimizations is that it overlaps multiple ORAM requests. 
\THISWORK\ supports overlapping ORAM requests rooted from LLC misses issued by different processes (such as concurrent KV access in \rediss\ or token feature table accesses in \llms\ among different users) for better resource availability in the cloud settings. 
While overlapping ORAM requests incur sharing of the ORAM controller and may interfere with other requests, we quantitatively show that \THISWORK\ by design ensures isolation of service latencies at the ORAM controller of each LLC miss. 
Fig.~\ref{fig:security_new_workloads} shows the analysis of ORAM response latencies for four workloads - \mcfs, \prs, \llms, and \rediss.
These figures clearly show that the access latencies are closely clustered together.
This is because ORAM requests are issued at a constant rate and once an ORAM request is issued at the ORAM controller, it always attempts to issue DRAM requests at its earliest time of resolving protocol dependencies and contend the memory controller with a fixed number of ORAM requests before its issue. 
Thus, the response latency of any ORAM request is a deterministic function of a constant number of uniformly random leaf selections before issue, leading to fully decoupled behavior of LLC misses, regardless of which process launched it. 
}

\textcolor{black}{
Note that all the access latencies cannot be identical due to the hardware architecture of memory controllers and DRAM.
However, the variance due to this does not render a design insecure as it is based only on the \textit{public information}, including the memory controller design, DRAM access protocol, and \texttt{PA} to ORAM leaf mapping (which is decoupled from \texttt{PA} as this is remapped randomly at each access).
For example, modern memory controllers reorder requests to exploit bank-level parallelism and row-buffer locality in DRAM.
This always leads to variations in DRAM response timings, and hence ORAM response timings.
\THISWORK\ design does not affect this behavior.
For the evaluated workloads, about 59\% of DRAM requests in \THISWORK\ access an already open row leading to lower latency, versus 37\% of accesses that require precharging a row and activating a new row leading to higher latency.
The similarity in these statistics for different workloads is owing to the ORAM protocol (similar to Fig.~\ref{figure:Ringoram_breakdown}(a)).
}

\textcolor{black}{
\begin{equation} 
\label{mutual_eq}
\begin{aligned}
\text{M}
&\equiv\frac{p_{1}}{2}log_2\frac{2p_{1}}{p_{1}+p_{2}}+\frac{p_{2}}{2}log_2\frac{2p_{2}}{p_{1}+p_{2}}& 
\\
&+\frac{1-p_{1}}{2}log_2\frac{2(1-p_{1})}{2-p_{1}-p_{2}}+\frac{1-p_{2}}{2}log_2\frac{2(1-p_{2})}{2-p_{1}-p_{2}}
\end{aligned}
\end{equation}
}

\textcolor{black}{
We further use mutual information M~\cite{goldsmith1997capacity, deng2019secure} between victim behavior B and attacker observation O to quantify the amount
of information about the response latency that an attacker can gain. 
From the potential attacker's sample of its access latency, we model the attacker's best guess about whether the victim's behavior hits the attacker's past access address is based on whether the access latency is lower/higher than the median observed latency. 
Using Table.~\ref{tab:bo_example}, Equation~\ref{mutual_eq}, and experimental measurements, the calculated mutual information \texttt{M} in these workloads is close to 0 (see table in Fig.~\ref{fig:security_new_workloads}).
M close to 0 indicates that p1, p2 $\approx$ 0.5: the attacker observes longer and shorter than median timings with almost 50-50 probabilities.
Therefore, the potential attacker cannot extract any information gain about the private program behavior out of \THISWORK.
}

\begin{table}
  \centering
  \scriptsize
  \caption{
\textcolor{black}{Probabilities of different victim behaviors B and attacker observations O.}}
  \begin{tabular}{|c|c|c|c|}
    \hline
    \multicolumn{2}{|c|}{} & \multicolumn{2}{|c|}{\makecell{Attacker's \\ observation \textit{O}}}  \\
    \cline{3-4}
    \multicolumn{2}{|c|}{} & Longer timing & Shorter timing \\
    \hline
    \multirow{2}{*}{\makecell{Victim's \\ behavior \\ B}} & \makecell{Requested block\\ is in the \texttt{Stash}} & $p_1$ & 1 - $p_1$ \\
    \cline{2-4}
    & \makecell{Requested block\\ is in the ORAM Tree} & $p_2$ & 1 - $p_2$ \\
    \hline
  \end{tabular}
  \label{tab:bo_example}
\end{table}

\section{Evaluation Methodology} \label{section:methodology}
\subsection{Real-World Cloud Services}
We use a variety of cloud services for evaluation as shown in Table~\ref{table:dataset}. 
The goal of ORAM is to hide the user's sensitive access patterns, such as accessed node IDs in graphs, item embedding IDs in deep-learning-based recommendation models (DLRMs), and token IDs in large language models (LLMs).
We measure up to 50M ORAM requests in the protected memory space and use the first half of the execution as a warmup, which translates into more than 3B memory instructions to the untrusted cloud memory.

\begin{table}[t]
\caption{Real-world services that demand obliviousness.} 
\centering 
\scriptsize
\begin{tabular}{c | c | c } 
\hline 
\textbf{Category} & \textbf{Workload Name} & 
\textbf{Description}
\\
\hline 
\hline 
\multirow{2}{*}{SPEC17} & \mcf\ (\mcfs)  & Route planning computation~\cite{bucek2018spec}
\\
& \lbm\ (\lbms) & Fluid dynamics computation~\cite{bucek2018spec}
\\
\hline
\multirow{2}{*}{Graph} & \pr\ (\prs) & Score ranking~\cite{beamer2015gap} on Livejournal~\cite{backstrom2006group}
\\
& \motif\ (\motifs) & Graph mining~\cite{mackey2018chronological} on Wikipedia~\cite{leskovec2010signed}
\\
\hline
\multirow{3}{*}{DL} & \rmone\ (\rmones) & Meta RM~\cite{gupta2020architectural, gupta2020deeprecsys} on Criteo 1T~\cite{criteo}
\\
&\rmtwo\ (\rmtwos) & Alibaba RM~\cite{zhou2018deep} on DBLP~\cite{rossi2015network}
\\
&\llm\ (\llms) & GPT-2~\cite{radford2019language} on OpenORCA~\cite{OpenOrca}\\
\hline
\multirow{3}{*}{KV} 
& \redis\ (\rediss) & Redis KV access~\cite{AmazonUserApi}
\\
&\stream\ (\streams) & Streaming memory access
\\
&\rdom\ (\rdoms) & Random memory access
\\

\hline 
\end{tabular}
\label{table:dataset} 
\end{table}

\subsection{State-of-the-art ORAM Baselines} \label{subsection:baselines}

\noindent
\textbf{PathORAM~\cite{pathoram}}
is a widely adopted protocol that achieves low algorithmic complexity and small on-chip requirements. 

\noindent
\textbf{RingORAM~\cite{ringoram}}
builds on top of PathORAM with significant bandwidth improvement. 

\noindent
\textbf{PageORAM~\cite{pageoram}}
uses PathORAM as the base protocol and introduces \textit{sibling node} accesses.
Sibling node accesses can expand the options for a block's residence and capitalize on the row buffer locality associated with accessing the nodes in PathORAM. 
The size of tree buckets can be reduced in this way, thereby reducing the overhead of each ORAM access.

\noindent
\textbf{IR-ORAM~\cite{iroram}}
uses hardware to keep track of the tree-top cache blocks' \texttt{PosMap} mappings. 
When IR-ORAM detects the accessed block hits in the tree-top cache, the accesses to the recursive \texttt{PosMap} ORAM are bypassed.
Additionally, IR-ORAM shrinks the bucket size in the middle part of the ORAM tree to reduce the overhead of each ORAM access.
IR-ORAM is similar and outperforms an early work Rho~\cite{rho}. 
We use IR-ORAM to represent this optimization idea.

\noindent
\textbf{PrORAM~\cite{proram}}
and its variants~\cite{che2020multi, rajat2023laoram} use prefetch to eliminate ORAM accesses (see \S\ref{section:prefetch_opt}). \textcolor{black}{The performance is measured after sweeping for best-performing prefetch lengths with Fat-Tree optimizations}~\cite{rajat2023laoram}.


\subsection{Simulation Parameters and Infrastructure} \label{subsection:simulation_infrastructure}
\begin{table}[t]
\caption{\THISWORK\ system configuration.}
\centering
\begin{footnotesize}
\scriptsize \begin{tabular}{ r || l }
\hline
\rule{0pt}{10pt}
\textbf{System Component} & \textbf{Modeled Parameters} \tabularnewline
    \hline
    \hline
\textcolor{black}{Host processor}
  &  \textcolor{black}{32-OoO cores, 4-wide issue, 2.66GHz frequency,}
  \tabularnewline
\textcolor{black}{}
  &  \textcolor{black}{128-entry ROB, 3-level inclusive cache hierarchy}
  \tabularnewline
\textcolor{black}{L1 cache}
  &  \textcolor{black}{32KB private per core, 4-way associative}
  \tabularnewline
\textcolor{black}{L2 cache}
  &  \textcolor{black}{256KB private per core, 8-way associative}
  \tabularnewline
\textcolor{black}{L3 cache}
  &  \textcolor{black}{8MB shared cache, 16-way associative}
  \tabularnewline
\hline
Protected memory space
    & 16GB user data protected \tabularnewline
Tree-top caches~\cite{maas2013phantom}
    & 24$\times$ banks of 32 KB scratchpad (3$\times$256 KB total)
    \tabularnewline
PosMap$^3$
    & 16$\times$ banks of 1 MB EDRAM (16 MB total)
    \tabularnewline
Stash$^0$, Stash$^1$, Stash$^2$
    & 3$\times$ cache bank of 16 KB SRAM cache (48 KB total)
    \tabularnewline
PE layout
    & 3 (row) $\times$ 8 (column) PEs, 1.6GHz frequency
    \tabularnewline
    \hline
Outsourced DRAM
  &  4-channel DDR4-3200, 102.4 GB/s peak bandwidth
  \tabularnewline
    \hline
\end{tabular}
\label{table:sysConfig1}
\end{footnotesize}
\end{table}
Table~\ref{table:sysConfig1} shows the modeled system configuration. 
Similar to prior works, \THISWORK\ uses 3 levels of sub-ORAM trees. 
The protected user memory space is 16GB. 
We use 256 KB tree-top caches, and 16KB stash for each level of sub-ORAM, the same cache size provision as prior works~\cite{iroram}.  


To accurately estimate the performance of \THISWORK, we implement a detailed two-phase simulation methodology.
First, we model all hardware components (except caches) using System Verilog HDL.
We synthesize this design using a commercial 28 nm technology library using the Synopsys Design Compiler.
Using detailed post-synthesis RTL simulations, we extract the critical path delay of our circuits and set \THISWORK\ clock frequency at 1.6 GHz.
Additionally, we collect the power and area numbers using RTL.
We use CACTI~\cite{cacti} to estimate the performance/power/area of SRAM caches.
Second, to model DRAM performance, \textcolor{black}{We utilize the widely-used, cycle-accurate Ramulator~\cite{ramulator} to model DRAM.
We additionally model the cycle-accurate behavior of the \THISWORK\ controller at the front end of the Ramulator, interacting with Ramulator memory events.}
To measure the effect on end-to-end performance, we use Sniper~\cite{carlson2011sniper}.
This simulator faithfully models all system components. 
We validate the simulator's functionality by comparing its memory traces with \THISWORK\ software version to ensure the absence of missed events.
\THISWORK\ is open-sourced at \textcolor{blue}{\url{https://github.com/Linestro/Palermo-ORAM}}. 

\section{Evaluation Results} \label{section:results}

\subsection{Performance Analysis}
\textbf{\THISWORK\ versus prior works.}  \label{perf:main}
\begin{figure*}
	\centering
		\includegraphics[width=0.98\textwidth, trim={0cm 0cm 0cm 0cm}, clip]{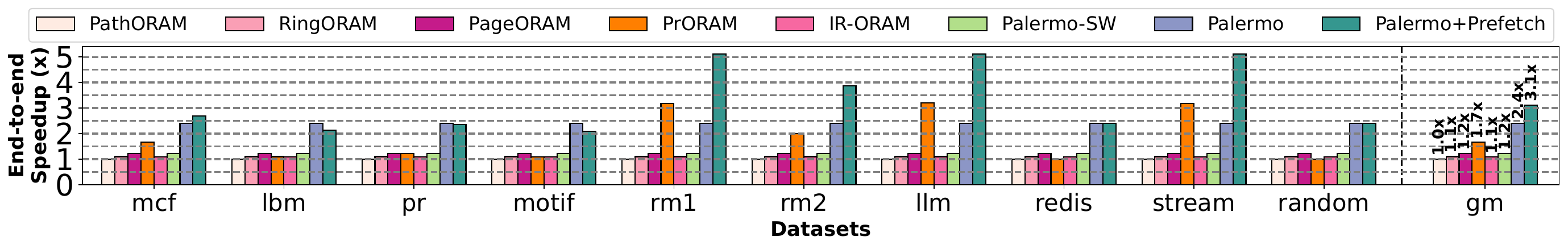}
		\caption{
        \textbf{End-to-end performance improvements while executing a variety of cloud service benchmarks with RingORAM~\cite{ringoram}, PageORAM~\cite{pageoram}, PrORAM~\cite{proram}, IR-ORAM~\cite{iroram}, \THISWORK\ software only version, and \THISWORK\ normalized to PathORAM performance. Higher is better. \textcolor{black}{PrORAM performance is measured after sweeping for the best-performing prefetch length with Fat-Tree optimizations. \THISWORK +Prefetch always applies the same prefetch length as PrORAM selects in each workload such that LLC miss traffics are the same. }}
        }
		\label{figure:main_result}
\end{figure*}
Fig.~\ref{figure:main_result} compares the end-to-end performance of \THISWORK\ with the state-of-the-art ORAM optimizations. 
~\textcolor{black}{In absolute numbers, \THISWORK\ issues and resolves at 3.8E6 LLC misses per second and RingORAM is at 1.7E6 LLC misses per second. 
The proportion of dummy ORAM requests becomes negligible after the warm-up phase. Given the fully DRAM-bound execution after applying ORAM, the DRAM-bound execution in nature diminishes the need for dummy requests when all benchmarks execute with ORAM. 
}
\THISWORK\ achieves an average of 3.1$\times$ and 2.4$\times$ speedup with and without prefetch over PathORAM while RingORAM, PageORAM and IR-ORAM achieve 1.1$\times$, 1.2$\times$, and 1.1$\times$.
PrORAM achieves superior performance by avoiding the ORAM protocol for a subset of memory requests when their data is prefetched and is present in LLC.
This is further evident by observing a stark performance difference in \streams\ and \rdoms, where PrORAM achieves suboptimal performance for workloads with little to no spatial locality. 
~\textcolor{black}{
\THISWORK\--SW shows the 1.2$\times$ protocol-level-only speedup over PathORAM with software mutex synchronizing the inter-request parallelism. 
\THISWORK\ shows an additional 2.6$\times$ performance brought by the co-designed hardware. 
The overall design aspects contribute a total of 3.1$\times$ to performance gain. 
}

\begin{figure}
	\centering
		\includegraphics[width=0.48\textwidth, trim={0cm 0cm 0cm 0cm}, clip]{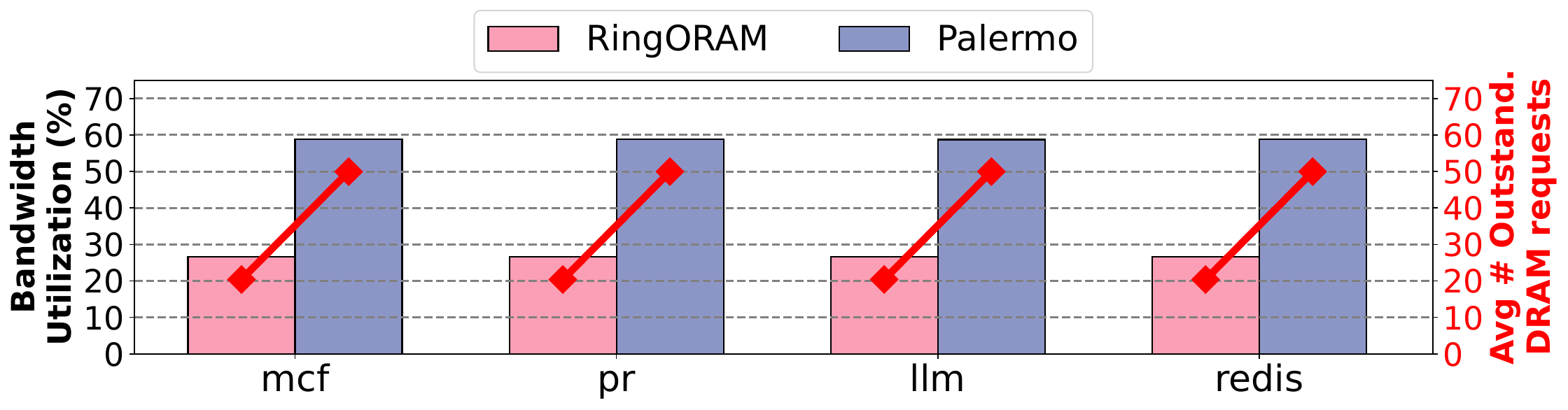}
		\caption{
        \textbf{DRAM bandwidth utilization comparison and average number of outstanding DRAM requests in RingORAM and \THISWORK\ without prefetch optimization.
        \THISWORK\ architecture improves parallelism and enqueues 2.8$\times$ more outstanding requests on average in the memory controller, resulting in a 2.2$\times$ higher DRAM bandwidth utilization.
        } 
        }
		\label{figure:explain}
\end{figure}

The performance improvements of \THISWORK\ are attributed to unlocking massive opportunities for parallelism.
The performance gains of \THISWORK\ can be further illustrated using the results in Fig.~\ref{figure:explain}.
This figure compares the bandwidth utilization and average number of outstanding DRAM requests in the memory controller.
\textcolor{black}{For better illustration, we show \THISWORK\ without prefetch such that the total DRAM traffic is identical between RingORAM and \THISWORK . }
Using \THISWORK\ protocol-hardware co-design, multiple requests can be issued to the memory controller as soon as their minimal dependencies are met.
Therefore, the average number of outstanding DRAM requests in \THISWORK\ increases by 2.8$\times$, which translates to a 2.2$\times$ improvement in the memory bandwidth utilization. 

\textcolor{black}{To compare the prefetch effect, we sweep the best performing prefetch length for PrORAM for each workload, and apply the same prefetch length for \THISWORK . 
This ensures the LLC miss traffics are the same after prefetch filtering between PrORAM and \THISWORK\ for a fair comparison. 
\THISWORK\ outperforms PrORAM by 1.9$\times$ due to better memory-level parallelism and no background eviction incurred by \texttt{Stash} pressure. 
}

\textcolor{black}{
Note that the bandwidth utilization remains consistent across various workloads. 
Our chosen workloads represent a wide range of popular applications and exhibit diverse memory behaviors in the absence of ORAM. 
The uniformity in bandwidth across these workloads stems from applying the ORAM protocol. 
Each memory address is mapped to a random leaf on the ORAM tree on every access. 
This consistent behavior on the memory bus illustrates how ORAM effectively obscures memory access patterns from potential attackers.
}

\label{perf:explanation}


\textbf{Bounded stash size in \THISWORK.}
\begin{figure}
	\centering
		\includegraphics[width=0.48\textwidth, trim={0cm 0cm 0cm 0cm}, clip]{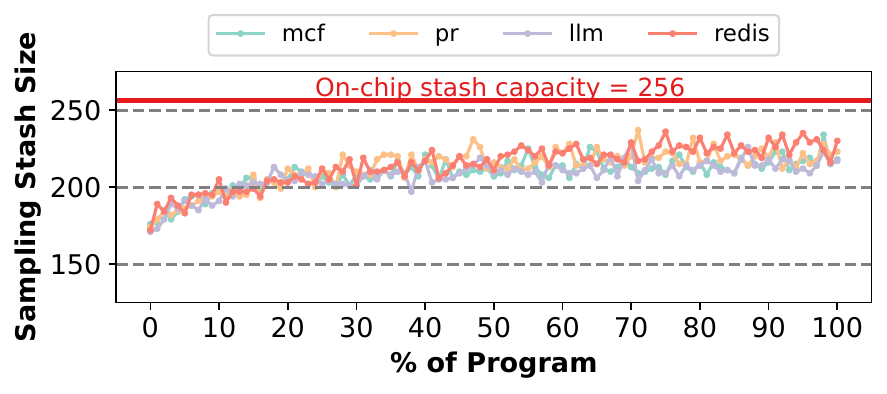}
		\caption{
        \textbf{\THISWORK\ stash utilization over time while executing various workloads.
        The empirical result shows that even with the introduction of concurrency with \THISWORK, the stash utilization is bounded.
        } 
        }
		\label{figure:stash}
\end{figure}
Stash overflows lead to background evictions, which adversely affect the performance of the ORAM protocol. 
A high-performance on-chip ORAM controller solution must achieve a negligible probability of overflowing even with a small stash (\textit{e.g.,} $< 2^{-103}$ in RingORAM with a 256-size stash).
One of the key design goals of \THISWORK\ is to improve performance with a bounded on-chip stash and avoid dummy requests.
Therefore, we study the occupancy of the stash in \THISWORK.
\THISWORK\ protocol synchronizes the \texttt{EP} phase after the \texttt{RP} phase to provide a theoretical bound of the stash size while concurrently serving requests.
Fig.~\ref{figure:stash} shows the sampled maximum stash size after every 1\% progress during the execution of \mcfs, \prs, \llms, and \rediss.
For the represented benchmarks, the maximum stash size throughout the program execution is 234, 237, 228, and 236.
This empirical result shows evidence that the intelligently designed \THISWORK\ protocol and hardware architecture maintain stash boundness and improve performance.

\subsection{\THISWORK\ Performance Analysis with Prefetching}
\begin{figure}
	\centering
		\includegraphics[width=0.48\textwidth, trim={0cm 0cm 0cm 0cm}, clip]{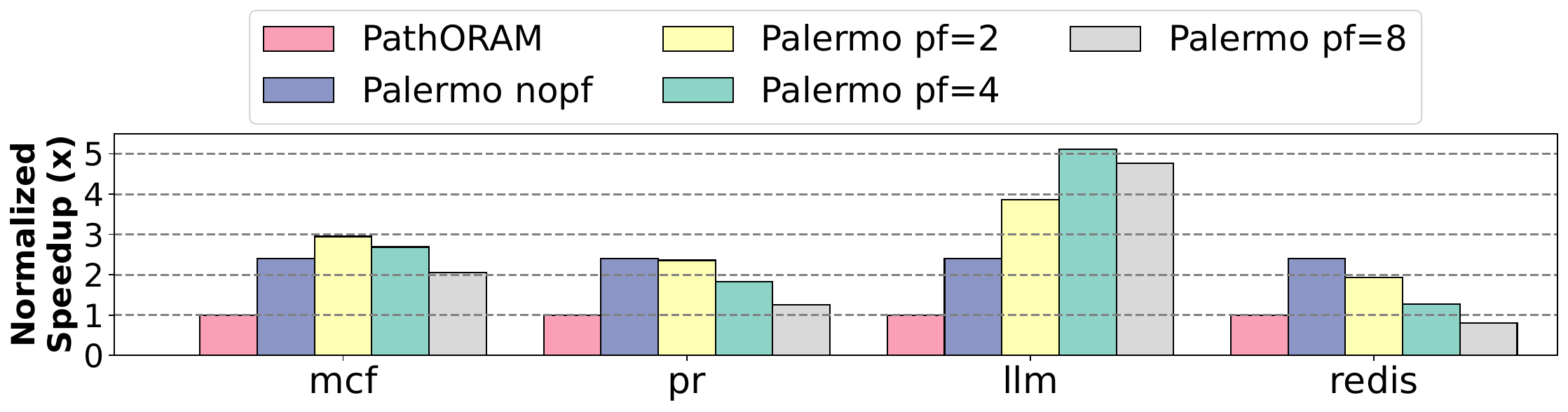}
		\caption{
        \textbf{
        Performance sensitivity of \THISWORK\ with different prefetch lengths. pf=X refers to converting \THISWORK\  \texttt{Data} ORAM tree block accesses in \texttt{RP} phase to a sequence of X DRAM accesses. 
        }}
		\label{figure:prefetch_sweep}
\end{figure}

\textcolor{black}{
Fig.~\ref{figure:prefetch_sweep} shows \THISWORK\ performance study of representative workloads with different prefetch lengths.
For SPEC, Graph, and KV access workload with moderate locality, \THISWORK\ performance only moderately changes and consistently outperforms PathORAM when switching prefetch length from 1 to 4. 
Thus, \THISWORK\ performance is not critically dependent on selecting the best prefetch length. 
For embedding access workloads, selecting prefetch length closing to the size of an embedding row offers maximized benefits.
When configured with prefetch length of X, \THISWORK\ \texttt{Stash} contains X cachelines for each SRAM entry. 
Thus, the amount of data in \texttt{Stash} grows by the dimension of X. 
However, the number of \texttt{Stash} tags remains bounded and stays below 256, regardless of the prefetch length. 
This is because \THISWORK\ prefetch strategy does not change the protocol algorithm of handling blocks but only magnifies the block size by a factor of X in \texttt{Data} ORAM tree. 
}
        
\subsection{Sensitivity Analysis}

\begin{figure}
	\centering
		\includegraphics[width=0.48\textwidth, trim={0cm 0cm 0cm 0cm}, clip]{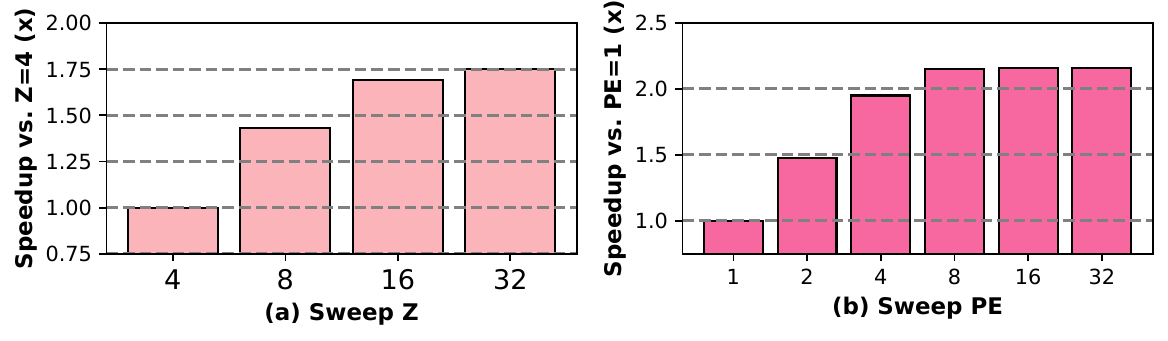}
		\caption{
        \textbf{\THISWORK\ performance sensitivity study with respect to (a) ORAM protocol parameter \texttt{Z}, and (b) the number of PE columns in \THISWORK\ architecture.} 
        }
		\label{figure:sweep}
\end{figure}

\textbf{ORAM Protocol Design Parameters. }
\THISWORK\ offers a rich protocol parameter sweep including the number of real and dummy blocks (\texttt{Z}, \texttt{S}) per \texttt{node} and eviction frequency \texttt{A}. 
We sweep all valid parameters shown in ~\cite{ringoram, cao2021streamline}. 
Fig.~\ref{figure:sweep}(a) shows the memory throughput of different protocol parameters. 
Interestingly, \THISWORK\ exploits more benefits from larger (\texttt{S}, \texttt{A}) because they create fewer write barriers for concurrent serving of multiple ORAM requests. 
With larger (\texttt{Z}, \texttt{S}, \texttt{A}) \THISWORK\ achieves up to 1.8$\times$ performance over (4, 5, 3) with the same capacity of protected memory space.
Larger \texttt{A} can create a higher pressure on \texttt{Stash} to store more blocks temporarily. 
\THISWORK\ adopts (16, 27, 20) configuration that can accommodate a modest stash size of 256 blocks.

\textbf{Number of PEs.}
To demonstrate the performance sensitivity of \THISWORK\ architecture, Fig.~\ref{figure:sweep}(b) shows the performance of executing \rdoms\ sweeping PE array size from 3$\times$1 to 3$\times$32. 
With fewer PEs, the workload is bound by structural hazards, instead of real dependencies between concurrent requests. 
Adding more PEs enables increased concurrency in issuing ORAM requests.
Our evaluation shows that with 3$\times$8 PEs, the workload starts to saturate the memory bandwidth and achieves 2.2$\times$ memory throughput over 3$\times$1 PEs.

\subsection{Area and Power Analysis} \label{perf:area}

\begin{figure}
	\centering
		\includegraphics[width=0.48\textwidth, trim={0cm 0cm 0cm 0cm}, clip]{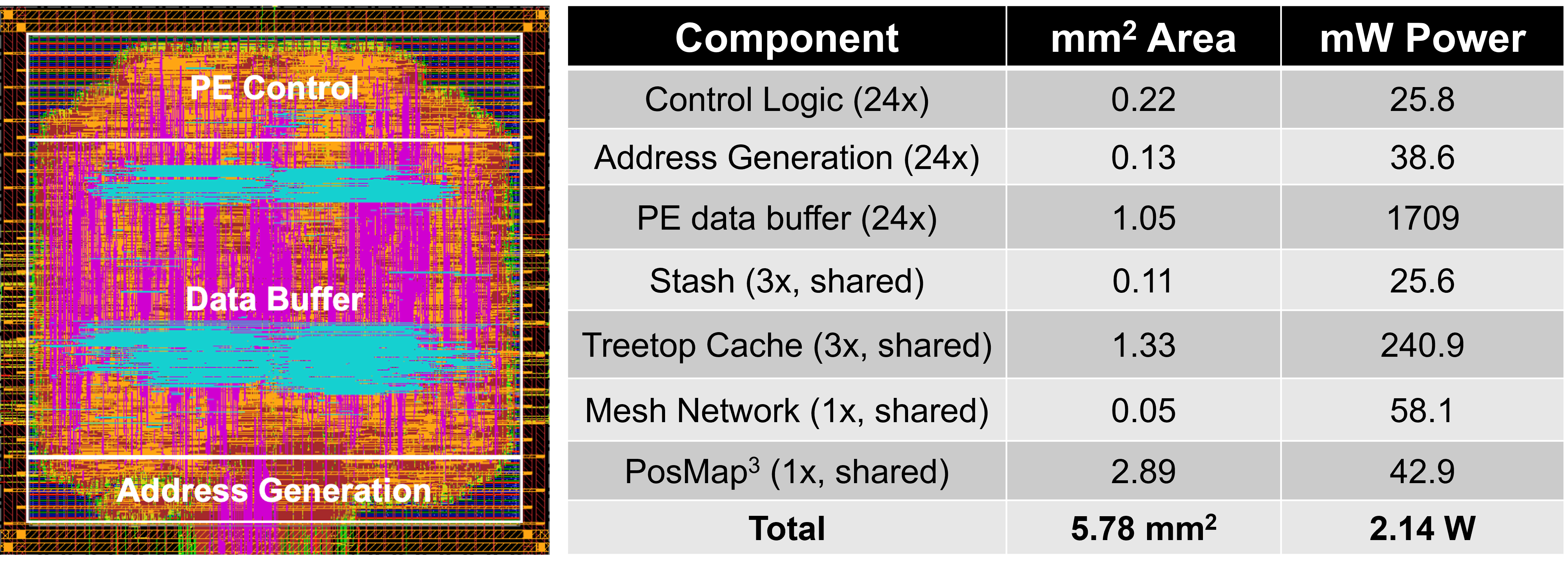}
		\caption{
        \textbf{A single PE layout, area, and power analysis of the entire \THISWORK. Power is measured at 1.6 GHz.}
        }
		\label{fig:area}
\end{figure}

Fig.~\ref{fig:area} shows the layout of a single PE, area, and power estimates for a full \THISWORK\ ORAM controller design.
The power results include both leakage and dynamic power consumption. 
The dynamic power is averaged over all workloads.
The table shows that \THISWORK\ consumes an area of 5.78mm$^2$ and 2.14W.
A majority of this is consumed in on-chip tree-top caches and PE data buffers.
A 2D pipeline of PE buffers is the key component of \THISWORK\ architecture to unlock memory-level parallelism by issuing concurrent request. 
The tree-top cache~\cite{maas2013phantom} stores the bottom level of the tree, which exhibits the highest access intensity.
Therefore, the share of their area and power is justified.
\textcolor{black}{
In comparison, the past FPGA work~\cite{maas2013phantom, fletcher2015low} Phantom operates at 200MHz and takes more than 20$mm^2$ using at least 5\% of the logic cells in high-performance Virtex-7 FPGAs. 
Other works optimizing ORAM~\cite{iroram, raoufi2023ab, rajat2023laoram} do not report area and power numbers.
To compare, \THISWORK\ offers a high-performance on-chip ORAM controller solution that serves LLC misses obliviously and interacts with DDR memory at 1.6GHz.
\THISWORK\ takes 5.78$mm^2$ area in 28nm technology and consumes 2.14W at 1.6GHz frequency.}

\section{Related Work} \label{section:related_work}

\textbf{Memory-level parallelism in ORAM.}
To enhance DRAM throughput of the ORAM protocol, prior works~\cite{cao2021streamline, zhang2015fork, fujieda2016last, che2019imbalance, zhang2018shadow, wang2017cooperative, omar2018breaking} implement on-chip and memory controller-level optimizations.
They result in sub-optimal speedup because they strictly serve ORAM requests one after the other.
\THISWORK\ addresses this aspect by proposing an optimized protocol and hardware architecture to overlap multiple ORAM requests and improve performance.

\textbf{ORAM capacity utilization optimizations.}
PathORAM uses 50\% of capacity to store dummy blocks~\cite{pathoram}, while RingORAM has an even higher dummy block percentage~\cite{ringoram}. 
AB-ORAM~\cite{raoufi2023ab} observes that RingORAM has many invalid blocks during the execution that can be recycled.
To recycle a block, AB-ORAM allocates blocks at a remote bucket that are marked as invalid.
Cao \textit{et al.}~\cite{cao2021streamline} applies "green blocks" that can reduce the block count in each bucket to save memory capacity.
\THISWORK\ is orthogonal to improving capacity.

\textbf{Other oblivious approaches.}
Private Information Retrieval (PIR)~\cite{lin2022inspire, mayberry2013efficient, lipmaa2010two, zhong2012pir} is an orthogonal approach to ORAM. 
PIR keeps data blocks static and hides the target of each query using costly homomorphic encryption.


\section{Conclusion} \label{section:conclusion}

To optimize ORAM performance, we introduced \THISWORK, a protocol and hardware co-design that enables the concurrent processing of multiple ORAM requests to maximize memory throughput while preserving ORAM correctness and security guarantees.
Using a diverse workload mix, we demonstrated that \THISWORK\ achieves 2.8$\times$ performance on average compared to a RingORAM baseline, with a negligible area overhead of 5.78mm$^2$ on a CPU, without compromising security.


\clearpage
\balance
\bibliographystyle{IEEEtranS}


\end{document}